# Ferroelectric quantum criticality and enhanced superconductivity in plastically deformed strontium titanate

S. Hameed[1,*], D. Pelc[2,1,*,#], Z. W. Anderson[1], A. Klein[3,1], R. J. Spieker[1], L. Yue[4], B. Das[5], J. Ramberger[5], M. Lukas[6], Y. Liu[7], M. J. Krogstad[8], R. Osborn[8], Y. Li[4], C. Leighton[5], R. M. Fernandes[1] and M. Greven[1,#]

[1]School of Physics and Astronomy, University of Minnesota, Minneapolis, MN, USA

[2]Department of Physics, Faculty of Science, University of Zagreb, Croatia

[3]Department of Physics, Faculty of Natural Sciences, Ariel University, Israel

[4]International Center for Quantum Materials, School of Physics, Peking University, Beijing, China

[5]Department of Chemical Engineering and Materials Science, University of Minnesota, Minneapolis, MN, USA

[6]Faculty of Mechanical Engineering and Naval Architecture, University of Zagreb, Croatia

[7]Neutron Scattering Division, Oak Ridge National Laboratory, Oak Ridge, TN, USA

[8]Materials Science Division, Argonne National Laboratory, Argonne, IL, USA

[*]these authors contributed equally

[#]correspondence to: dpelc@umn.edu, greven@umn.edu

**The properties of quantum materials are commonly tuned using experimental variables such as pressure, magnetic field and doping. Here we explore a different approach: irreversible, plastic deformation of single crystals. We show for the superconductor $SrTiO_3$ that compressive plastic deformation induces low-dimensional superconductivity significantly above the superconducting transition temperature ($T_c$) of undeformed samples, with evidence of superconducting correlations at temperatures two orders of magnitude above the bulk $T_c$. The superconductivity enhancement is correlated with the appearance of self-organized dislocation structures, as revealed by diffuse neutron and X-ray scattering. We also observe signatures of deformation-induced quantum-critical ferroelectric fluctuations**

**and inhomogeneous ferroelectric order via Raman scattering. These results suggest that the strain surrounding the self-organized dislocation structures induces local ferroelectricity and quantum-critical dynamics that strongly influence $T_c$, consistent with a theory of superconductivity enhanced by soft polar fluctuations. More broadly, our results demonstrate the promise of plastic deformation and dislocation engineering as tools to manipulate electronic properties of quantum materials.**

**Introduction.** The application of reversible, elastic uniaxial strain has emerged as a powerful means to study and manipulate quantum materials[1-4]. The effects of uniaxial stress *beyond* the elastic regime, however, are not widely studied in single-crystalline materials. The resultant plastic deformation fundamentally differs from elastic strain, as it creates extended defects – typically dislocations – and can induce their self-organization into structures spanning broad length scales[5]. The local atomic arrangement near dislocation cores is very different from the bulk and significantly modifies nearby electronic properties[6-9]. Such effects are amplified when dislocations assemble into larger-scale structures[10].

Here we use electron-doped strontium titanate, $SrTiO_3$ (STO), as a model system to study the effects of significant plastic deformation on electronic properties. We show that superconductivity in STO is considerably enhanced by plastic deformation, and that self-organized dislocation structures play a crucial role. Although superconductivity in the cubic perovskite STO has been known for decades[11], its origin remains poorly understood and cannot be explained by the standard electron-phonon pairing mechanism, as it prevails to extremely low charge-carrier densities, where the Debye frequency is much larger than the Fermi energy[12-14]. Like in other unconventional materials, superconductivity appears in a dome-shaped region of the temperature-doping phase diagram. Moreover, STO is based on the perovskite crystal structure, similar to other superconducting materials of high interest, including bismuthates, ruthenates, and high-$T_c$ cuprates[15].

Perhaps less widely appreciated, STO also has remarkable mechanical properties for a ceramic material. Single crystals can be plastically deformed to a high degree in compression, without macroscopic cracking, even at ambient temperature[16]. Although the microscopic origin of the extended plastic regime in STO is debated[17], it is known that plastic deformation occurs via the creation, dissociation and glide of dislocations in {110} planes of the cubic structure[16,17]. At elevated stresses, these dislocations coalesce into bundles, which leads to work hardening[16] (see Figs. 1a, S1). We find that atomic substitution (at least at the levels relevant to electronic doping) does not significantly influence ductility. STO therefore provides an ideal opportunity to investigate effects of extensive plastic deformation on the electronic properties of a model quantum material.

**Structure and dynamics of plastically deformed STO.** Using specially designed uniaxial pressure cells (Fig. 1a), we have routinely been able to deform STO crystals up to engineering strains ε ~ 8% in compression at 300 K (*e.g.*, Fig. 1a, Fig. S1). Macroscopically, deformation induces streaks at the crystal surfaces, perpendicular to the stress (Fig. S2), indicative of a highly inhomogeneous strain distribution on the 1-10 μm scale. To further investigate the structural effects of deformation, we performed neutron and X-ray diffuse scattering measurements. As expected, undeformed samples show sharp Bragg peaks (Fig. S3). In contrast, deformation induces pronounced azimuthal elongation of the Bragg peaks in planes perpendicular to the surface streaks (Fig. 1b), but not in other planes (Fig. S3d). Similar features, known as asterisms, are observed in deformed metals[18] and are due to tilts of mesoscopic, relatively defect-poor crystalline regions bounded by dislocation structures,[19,20] as illustrated schematically in Fig. 1c. Importantly, asterisms only appear above ~0.5% deformation (Fig. S4), with little associated change in dislocation density (Fig. 1d). Also, deformation does not considerably influence the low-temperature tetragonal structure[21]; the usual tetragonal superstructure peaks are observed at low temperatures, and exhibit asterisms.

Away from the Bragg peaks, we observe features that are not present in undeformed samples, most prominently cross-like diffuse patterns (Figs. 2a,b, S5). The orientation of the streaks in these patterns is consistent with the presence of {110} slip planes, and their small width and grainy substructure indicates long-range dislocation correlations. This, along with the presence of asterisms, implies that the diffuse scattering originates from dislocations that self-organize into planar structures – walls – that act as low-angle tilt boundaries (Fig. 1c, 2e). We obtain good agreement with experiment by calculating the scattering pattern associated with the strain field of dislocation walls with {110} slip planes (Fig. 2c,d). The strain fields from such remarkable self-organization of dislocations play an important role in determining local electronic properties, *e.g.*, ferroelectricity[4,22,23,24]. Recent piezo-force-microscopy and nonlinear optical investigations indeed show that individual dislocation structures have local polarization[8,25], while the polar displacements are not significantly affected by electron doping[26]. Furthermore, modeling suggests that local electric polarization is enhanced in extended dislocation structures[10].

We find clear evidence of deformation-induced ferroelectricity in Raman scattering experiments on undoped and Nb-doped samples (Fig. 3a-d). Two main effects are observed in deformed samples: the appearance of hard transverse optical (TO) modes related to inversion-symmetry breaking (see also Fig. S7), and a large, broad feature at low energies that increases sharply upon cooling. The presence of the TO2 and TO4 polar modes confirms that inversion symmetry is locally broken. Furthermore, the low-energy spectral-weight build-up is a direct signature of quantum-critical ferroelectric fluctuations that are present in both undoped and doped samples. This is most clearly seen in the real part of the static Raman susceptibility $\chi'(0)$ (Fig. 3e), which is obtained from the measured frequency-dependent imaginary-part $\chi''(\omega)$ *via* Kramers-Kronig transformation, analogous to, *e.g.*, studies of nematic fluctuations in iron-based superconductors[27] (see also Methods). The susceptibility $\chi'(0)$ follows a simple Curie-Weiss temperature dependence that does not saturate at the lowest measured temperatures, indicative of the quantum-critical nature of the fluctuations. Moreover, the dynamic susceptibility shows a broad peak that represents a characteristic fluctuation energy scale that softens on cooling (Fig. 3a-d,f). This signal is the well-known softening of the TO1 mode in STO. The dislocation-induced Raman signature is thus direct evidence of quantum-critical dynamic fluctuations in STO, accompanied by local ferroelectric order. Importantly, these fluctuations are only accessible to the inversion-symmetric Raman probe due to the presence of locally-broken inversion symmetry, which is a direct consequence of deformation-induced dislocation structures and inhomogeneity. In contrast, the TO1 elastic scattering in STO remains narrow down to low temperatures[28]. This highlights the fact that deformed STO is fundamentally different from the uniformly strained material. As we argue below, the Raman results imply the existence of regions with dynamic ferroelectric fluctuations that are enhanced beyond the bulk value and lead to local superconductivity enhancement.

**Enhancement of superconductivity.** Figure 4 shows low-temperature resistivity and magnetic susceptibility data for several deformation and Nb doping levels. A strongly anisotropic superconducting state appears in deformed samples, with typical midpoint transition temperatures of 0.6-0.7 K, well above the optimal resistive $T_c$ in undoped Nb-doped bulk STO[12,13,29] (Fig. 4a). For a sample with 0.2% Nb, for example, the Hall electron density is about $3·10^{19}$ cm$^{-3}$, the bulk $T_c$ without deformation is <0.35 K (Fig. S8), and a 5-6% deformation induces a $T_c$ enhancement of roughly a factor of two (Fig. 4a,b). The resistive transition is first observed in the resistance perpendicular to the stress ($R_\perp$), whereas in the parallel direction, the resistance ($R_\parallel$) first increases before decreasing again on cooling (Fig. 4b-d). Similar resistivity peaks have been observed around $T_c$ in doped STO films upon application of magnetic fields and at EuO/KTaO$_3$ interfaces[30,31]. Such anisotropic resistivity is expected in an inhomogeneous superconductor when resistances are measured in a van der Pauw geometry[32]. In our case, the lower resistivity is always $R_\perp$, which, when taken with the scattering data, indicates low-dimensional transport along the dislocation structures. In some samples, two or three distinct resistivity steps can be discerned (e.g., Fig. 4c), also consistent with inhomogeneous superconductivity. In samples with 0.2% Nb doping, which we studied at several deformation levels (Fig. S8), the superconductivity enhancement roughly correlates with the appearance of asterisms in diffuse scattering (Fig. S4). This suggests that the self-organized dislocation structures play a major role in the $T_c$ enhancement. The decrease in the perpendicular resistivity channel matches well with signatures in the diamagnetic response and with the onset of nonzero nonlinear susceptibility (Fig. 4f), leaving no doubt that this is a superconducting transition. The diamagnetic response is also strongly anisotropic, and the critical fields differ along and perpendicular to the stress direction (Fig. 4e). We also measured several deformed samples that were oxygen-vacancy-doped (OVD), with Hall electron concentrations up to $10^{19}$ cm$^{-3}$; no resistive transition was detected down to 360 mK.

In addition to the clear enhancement of superconductivity below 1 K, deformed STO samples show intriguing transport and magnetic properties at much higher temperatures. First, we note that a small diamagnetic response is present in the ac susceptibility up to at least 1.5 K (Fig. 4f). Furthermore, resistivity anisotropy develops already at around 30-50 K, and $R_\perp$ falls below the robust normal-state $T^2$ dependence characteristic of undeformed samples[33]. This is observed in both OVD and Nb-doped samples (Figs. 5a,b, S9, S10), but is more pronounced in OVD crystals with low electron densities. There, we observe a resistivity anisotropy of two orders of magnitude (Fig. 5a, S7), with $R_\perp$ roughly one order of magnitude below $R_0$, the value in undeformed samples. The difference is even larger if we consider the increase of the normal-state residual resistivity due to deformation (Fig. S9). We see no significant dependence of the resistivity downturn on the type of dopant, since the downturns are similar in OVD and Nb-doped samples (Fig. S10). Along with the resistivity decrease, we find significant nonlinear resistivity in the same temperature range, as well as nonlinear current-voltage characteristics at low temperatures (Fig. 5c). Most likely, these effects do not result from sample heating (see Methods).

We expect the deformation effect on the resistivity to consist of an increase due to added dislocations and a decrease related to superconducting correlations. To disentangle these contributions, we extract, for a set of 0.2% Nb-doped samples, both the measured residual resistivity ratio (RRR) and an RRR determined by fitting the high-temperature resistivity data to a $T^2$ form (Fig. 5d, inset). The increase of the extrapolated residual resistivities with strain (Fig. 5d) is consistent with a factor of ~2 increase of dislocation density between 0% and 1% deformation[34] (Fig. 1d); it is also isotropic (see Supplementary Material). However, comparison of measured and extrapolated RRR values indicates a roughly threefold reduction of the resistivity (Fig. 5d). In addition, applied magnetic fields increase the low-temperature resistivity more than expected from previously established magnetoresistance values[29,35] (Fig. S11), while the carrier concentration is unaffected by deformation: the Hall number shows little temperature dependence and is roughly consistent with the carrier density expected from the Nb concentration[29,35] (Fig. S12). Importantly, the Hall coefficient does not depend on the current direction, which is strong evidence that the deformation does not induce carrier-density changes or anisotropic dislocation scattering that might affect the resistivity anisotropy.

Our analysis suggests that superconducting correlations appear in deformed STO at temperatures as high as 30-50 K. We attempted to search for a corresponding diamagnetic signal, without conclusive results. The main issue is the presence of weak defect-induced magnetism[36] in both deformed and undeformed samples. In any case, if the superconductivity is low-dimensional, as might be inferred from the strong resistance anisotropy and high critical field scale, no appreciable diamagnetic shielding would be expected[37]. We note that the emergence of high-temperature superconducting correlations is not the only possible explanation for the observed behaviour, which could be the result of, e.g., shielding of point defects within regions of large dislocation densities, strain-induced hardening of phonon modes, or the formation of a low-dimensional ordered state (such as a charge-density wave) on dislocation bundles. Undoped STO also enters the quantum paraelectric regime[12] between 30 and 40 K, which could influence transport and magnetic properties in the presence of dislocation structures. However, the presence of quantum-critical fluctuations should lead to a resistivity increase due to the large available phase space for low-frequency scattering, except if the fluctuations lead to superconducting pairing. We also note that possible signatures of superconductivity at ~5 K were previously observed in undoped STO subjected to dielectric breakdown[38], which might also induce extended defects. Further experiments focusing on local properties, such as the local density of states or the local diamagnetism, are needed to firmly establish the presence of higher $T_c$ superconducting regions.

**Ferroelectric fluctuations and superconductivity.** We now propose a simple scenario to connect the existence of self-organized dislocation structures observed by diffuse scattering, the onset of inhomogeneous ferroelectric order and enhanced ferroelectric fluctuations observed by Raman scattering, and the local enhancement of $T_c$ observed by transport measurements. STO is a quantum paraelectric with an anomalously high dielectric constant thought to result from fluctuations of a soft polar TO phonon. Application of strain or "negative" pressure[39,40] can further soften the TO mode, tuning STO across a ferroelectric quantum critical point (QCP)[4,22]. Although the precise mechanism is debated, standard arguments imply that the soft-phonon contribution to $T_c$ is enhanced near the QCP[41,42,43,44], as indeed observed experimentally[1,26,45]. Since dislocations induce long-range strain fields, their presence induces a local modification of the TO mode frequency. This modification has two effects. First, the *strong* local strain at the dislocation structures induces anisotropic ferroelectric order in their vicinity at temperatures far above $T_c$. The static order breaks inversion symmetry, and therefore the inversion-even Raman probe directly couples to the inversion-odd ferroelectric fluctuations. Second, due to the long-range strain decay, there necessarily exist regions in which the TO-mode frequency is suppressed to zero, at least within a mean-field picture. If we assume that $T_c$ increases with the softening of the TO mode, such regions will show a local $T_c$ enhancement. We analyze this effect via the Ginzburg-Landau free energy of the TO mode coupled to the strain field, which we assume to vary slowly over distances **r** much larger than a lattice constant,

$$F(\mathbf{r}) \propto (\omega_T^2 + c_T^2 q^2)(\delta_{ij} - \hat{q}_i \hat{q}_j) u_i(\mathbf{r}, \mathbf{q}) u_j(\mathbf{r}, \mathbf{q}) + g_{ijkl} u_i(\mathbf{r}, \mathbf{q}) u_j(\mathbf{r}, \mathbf{q}) \varepsilon_{kl}(\mathbf{r}) + \cdots \quad (1)$$

Here, $u_i$ are components of the TO displacement field with wavevector **q**, $\omega_T^2 \approx (2meV)^2$ is the soft transverse phonon frequency, $\delta_{ij} - \hat{q}_i \hat{q}_j$ projects onto the transverse component, $\varepsilon_{ij}(\mathbf{r})$ is the slowly varying strain field generated by the dislocation structure, $g_{ijkl}$ is a coupling term, and we sum over repeated indices[46] (see Methods). Cubic symmetry is assumed for simplicity. The effect of the strain field is twofold: it splits the triply degenerate $\omega_T$ mode, and makes it inhomogeneous. Close to the dislocation core, the generated strain field, which can be calculated using standard elasticity theory (see Methods), is very large and induces ferroelectric order. While Eq. (1) is not justified in this regime, as it neglects dynamics and higher-order terms, the ferroelectric order near the core is nevertheless expected to suppress superconductivity[26]. Farther from the core, where Eq. (1) is valid, we find wide regions where the TO mode is significantly softened. In Fig. 6a, we plot the local dimensionless quantity $r_{min} = \omega_{min}^2/\omega_T^2$, where $\omega_{min}$ denotes the smallest TO mode, around a single dislocation core. This quantity measures the distance to the ferroelectric transition, with $r_{min} > 0$ denoting the disordered phase ($r_{min} = 1$ being the value for the undeformed sample) and $r_{min} < 0$ the ferroelectric state (where $\omega_{min}$ is imaginary). The shaded region is the area, of order $10^6 a^2$ with $a$ denoting the lattice constant, where $r_{min}$ is suppressed by a factor of 2 or more with respect to its value in the undeformed sample (see Methods). For the experimentally relevant case of the dislocation wall shown in Fig. 2e, interference cancels out the long-range strain and leaves an enhancement region of area $A \sim La$, where $L$ is the length of the dislocation wall, as shown in Fig. 6b. In the three-dimensional material, this structure results in a sheet-like region with suppressed $r_{min}$. While the functional dependence of $T_c$ on $r_{min}$ is complicated and not known near vanishing frequencies, it is well established that $T_c$ increases with decreasing $r_{min}$ in an exponential fashion not too close to the QCP[41,42,43]. Our calculations therefore demonstrate the tremendous potential for targeted strain-engineering of superconductivity in STO and other materials.

**Discussion.** Our work shows that plastic-deformation-induced dislocation structures enhance ferroelectric fluctuations and superconductivity in STO. This has wide-ranging implications for both the superconducting pairing mechanism in STO and similar materials, as well as for diverse material systems where plastic deformation, and dislocation engineering more generally, could be used to tune local electronic ordering. Superconducting pairing in STO has been attributed to electron-phonon interactions, plasmons, ferroelectric quantum critical fluctuations, local modes, and oxygen-vacancy negative-$U$ centres[13]. All of these might be strongly affected by plastic deformation. Even for undeformed STO it was recently found that intrinsic structural inhomogeneity influences superconducting correlations[47]. Moreover, tetragonal structural domain walls, which might be affected by deformation (although we do not see evidence for that in diffuse scattering), may also be important for pairing[13,48]. We also note that there is a precedent for pairing at domain walls[37] in $WO_3$. However, our calculations, which demonstrate that $T_c$ is enhanced by incipient strain-induced ferroelectricity, suggest that this mechanism plays a pivotal role in the effects observed here. Although our Raman scattering results provide unequivocal evidence for deformation-induced inhomogeneous fluctuations, further studies of local modes and phonon spectrum modifications in deformed STO and related quantum paraelectrics will be of great interest. Furthermore, considering that the Fermi-liquid-like normal-state $T^2$ resistivity itself is not understood, more detailed studies could help unravel the physics behind this puzzling phenomenon[33,49].

More broadly, plastic deformation and dislocation engineering can be used to manipulate any properties of quantum materials that couple to strain. A necessary prerequisite for this is to reach the plastic deformation regime; STO is a model system because of its remarkable ambient-temperature ductility, but most materials become ductile close to their melting temperature or under high strain rates. In particular, we expect that dramatic superconductivity enhancement is possible in other materials that are close to a ferroelectric instability, such as $PbTe$[50] and $KTaO_3$[31,51], as well as other nearly cubic perovskites like the high-$T_c$ bismuthates[52]. Cuprate high-$T_c$ superconductors also share important structural and superconducting features with STO[47], and their $T_c$ changes rapidly with strain[53]. Dislocation engineering could thus be a route to significantly enhance cuprate superconductivity. Dislocations have also recently been suggested to play an important role in the unconventional superconductor strontium ruthenate[9]. Other electronic ordering tendencies are influenced by dislocations as well; e.g., ferromagnetic dislocations have been found[6] in antiferromagnetic NiO, and multiferroicity has been predicted[7] for dislocations in ferroelectric $PbTiO_3$. Dislocation engineering could therefore lead to a broad range of novel functional materials.

**Methods**

*Samples.* Single crystals of undoped and Nb-doped STO were sourced commercially (MTI Corp.) and polished to have parallel and perpendicular faces for uniaxial pressure experiments. Oxygen-vacancy doping (OVD) of Nb-free STO was achieved through annealing in a high vacuum of $10^{-5}$ mbar with a titanium getter at 800°C for 2 h. Samples for transport measurements were 0.5 mm thick and typically ~1-2 mm laterally, while the neutron scattering samples were 1 mm thick and 5-6 mm laterally.

*Uniaxial pressure.* Two different custom-built uniaxial pressure cells were used to deform STO single crystals for transport, magnetometry and neutron scattering measurements. Both cells use the same operating principle: the force on the sample is supplied by the piston of a pneumatic cylinder, and the sample deformation is measured using a linear variable transformer (i.e., a set of two coaxial coils). Such a design is scalable and enables the application of very high forces, while stress-strain diagrams can be easily measured, since the displacement is determined independently. The pressure cell used for smaller samples has a maximum force of 350 N and uses two coaxial sapphire rods as anvils for sample compression. Our large sample cell has a maximum force of 5000 N, and sintered carbide plates are used as anvils. Both systems are operated with the same high-precision pressure regulator, with a maximum operating gas pressure of 10 bar. The high ductility of STO at ambient temperature makes it very convenient for plastic deformation studies, since it is not necessary to perform the deformation at high temperatures. The sample is always compressed along the [010] direction. Out of the six possible {110} slip planes, four are found to be active under plastic deformation for the applied stress direction. However, experimentally, we find that only two out of these four slip planes are active, as shown in Fig. 2e, most likely due to sample geometry.

*Diffuse neutron scattering.* These experiments were performed on the CORELLI[54] spectrometer of the Spallation Neutron Source at Oak Ridge National Laboratory, USA, in a closed-cycle refrigerator with a base temperature of 6 K. The sample masses were ~300 mg, with counting times of 9 h per temperature. The Mantid package was used for data reduction, including Lorentz and spectral corrections[55]. Several crystals with different deformation levels were annealed *ex situ* in vacuum under the same conditions as the transport samples.

*Diffuse X-ray scattering.* The experiments were performed on Beamline 6ID-D of the Advanced Photon Source, Argonne National Laboratory, USA. The beamline employs a superconducting magnet undulator and double-crystal monochromator to produce a high-energy monochromatic X-ray beam. For our experiments, the X-ray energy was 87 keV. A Pilatus 2M CdTe detector, whose sensor layer is optimized for high energies, was used to collect frames at 0.1 s exposure while samples were continuously rotated at 1°/s about a horizontal axis. Three sets of rotations were performed in each measurement between which (i) the detector was translated by 5mm in both the horizontal and vertical directions in order to cover gaps in the scattering between detector chips, and (ii) the sample rotation axis was offset by ±15° from perpendicular to the beam in order to allow artifacts caused by scattering in the detector sensor layer to be masked. The counting time at each temperature setting was 20 minutes. See the Supplementary Material in [56] for more details. The data covering a range of ~±15 Å$^{-1}$ in all directions, were collected down to 30 K using an Oxford N-Helix helium cryocooler, and transformed to $S(\boldsymbol{Q})$ using the software package CCTW (https://sourceforge.net/projects/cctw/).

*Raman scattering.* For Raman scattering measurements we used a Horiba Jobin Yvon LabRAM HR Evolution spectrometer with a liquid-nitrogen-cooled CCD detector and 1800 gr/mm grating. The measurements were done in the confocal backscattering geometry. The BragGrate notch filters were used to acquire data at low energies, down to around 10 cm$^{-1}$. The He-Ne laser line wavelength was 632.8 nm. The sample was held in a liquid-helium-flow cryostat. The focused laser beam spot was about 10-µm-diameter on the sample surface.

To obtain the imaginary part of the Raman susceptibility, $\chi''(\omega)$, the raw intensities were divided by the Bose factor. The static real part is calculated from the frequency-dependent imaginary part using the Kramers-Kronig relation

$$\chi'(0) = \frac{2}{\pi} \int_0^\infty d\omega \chi''(\omega)/\omega$$

To obtain the integrals, the imaginary part of the Raman conductivity $\chi''(\omega)/\omega$ was fit to a Lorentzian curve up to 140 cm$^{-1}$ and the area under the curve was calculated after subtraction of a constant offset.

*Transport and magnetization measurements.* For low-temperature measurements we used a home-built $^3$He evaporation refrigerator with external gas handling and a commercial sorption-pumped system. AC susceptibility was measured with the sample mounted in a conventional three-coil system, using a lock-in amplifier, both to provide the excitation current and to measure the induced voltages. We typically used excitation field amplitudes of ~1 Oe, frequencies of ~2 kHz, and a transformer preamplifier. Third-order magnetization was measured by detecting the induced voltage at the third harmonic of the fundamental excitation frequency, similar to previous work[47]. Transport measurements were performed using a high-throughput dipstick probe that was directly inserted into the $^3$He pot in the case of the home-built $^3$He system and in vacuum for the commercial $^3$He system. We used a van der Pauw contact geometry, ac excitation currents of 1 mA RMS at 13.4 Hz, with the signal detected using an AC resistance bridge or a lock-in equipped with a low-noise bipolar preamplifier. The temperature dependence of the measured voltages did not depend on precise contact placement, indicating that the samples are macroscopically homogeneous. For transport measurements above 2 K, we used a Quantum Design, Inc. MPMS system with a home-built AC resistivity probe employing a balanced source and lock-in amplifier, and a Quantum Design, Inc. PPMS, where DC currents (and current reversal) were used. Currents were typically between 1 mA and 10 mA, with higher currents used for samples with lower resistances. Contacts were made by wire-bonding aluminium or gold wires to sputtered gold pads, or by bonding Al wire directly to the sample surface. Contact resistances were typically a few Ohms at ambient temperature, and no nonlinearities were seen in our temperature and current range (except the intrinsic low-temperature effects, Fig. 5c). For measurements of the third-order nonlinear response in OVD samples, we employed a current of 2 mA RMS at 827 Hz. Voltage was detected at the third-harmonic frequency of the fundamental, and the phase was adjusted to three times the first harmonic phase to obtain correct separation of real and imaginary parts[57]. With this procedure, the imaginary part was negligible over the entire temperature range, as expected for nonlinear resistance. The sample resistance was below 10 mΩ at the relevant temperatures, which led to a heating power of less than 1 µW even at the highest applied currents. More importantly, we employed the high-frequency current for the measurement of third-order nonlinear response to eliminate sample temperature changes within one current oscillation. The Hall voltage was measured with a Hall bar contact configuration using the 7 T MPMS and 9 T PPMS magnets, with field and current reversal (and AC current in the MPMS). From the dependence of the Hall voltage on thickness and contact placement/size, we estimate that the corrections due to geometry and contact shorting are below 20%.

*Theory of the local ferroelectricity, the $T_c$ enhancement and the diffuse scattering pattern.* For simplicity, we assume a cubic system and neglect lattice distortions and the antiferrodistortive transition in STO. The susceptibility of a polar optical mode can be written as[42,43]

$$\frac{\chi_{ij}^{-1}(\mathbf{q},\Omega)}{E_T} = -\Omega^2 \delta_{ij} + (\omega_L^2 + c_L^2 q^2)\hat{q}_i\hat{q}_j + (\omega_T^2 + c_T^2 q^2)(\delta_{ij} - \hat{q}_i\hat{q}_j) + \Lambda_{ij} \quad (2)$$

Here, $\omega_T$ and $c_T$ ($\omega_L$ and $c_L$) denote the frequency and velocity of the transverse (longitudinal) component of the mode, and $E_T$ is an unimportant overall scale. We assume that $\omega_T \ll \omega_L$ so that the longitudinal mode can be safely integrated out. The coupling to the strain is given by $\Lambda_{ij} = \lambda_1(3\varepsilon_0)\delta_{ij} + \frac{3}{2}\lambda_2(\varepsilon_{ii}\delta_{ij} - \varepsilon_0\delta_{ij}) + \lambda_3(\varepsilon_{ij} - \varepsilon_{ii}\delta_{ij})$, where the strain tensor elements have been decomposed in the $A_{1g}, E_g$ and $T_{2g}$ irreducible representations of the cubic group, with $\varepsilon_0 = \frac{1}{3}\sum_{i=1..3}\varepsilon_{ii}$ denoting a volume change.

To simplify what follows, we neglect the effects of $\Lambda_{ij}$ on the irrelevant longitudinal mode and set the coupling to the shear strain to zero (i.e., $\lambda_3 = 0$). Then, the matrix $\Lambda_{ij}$ is diagonal, and the effect of the strain field is to lift the degeneracy of the modes in the transverse sector $\omega_T^2\delta_{ij} \rightarrow \omega_i^2\delta_{ij} = (\omega_T^2 + \Lambda_{ii})\delta_{ij}$. To find the strain field we solve the textbook problem of an edge dislocation along the c-axis[58]. In complex notation, where $u = u_x + iu_y$ is the displacement and $b = b_x + ib_y$ is the Burgers vector in the plane normal to the c-axis, the solution for a dislocation at position $z_0 = x_0 + iy_0$ is, up to an unimportant constant, given by $u(b, z - z_0)$, where

$$u(b,z) = \frac{-ib}{2\pi}\log z + \frac{1}{4(1-\sigma)}\left(\frac{ib}{2\pi}\log z\bar{z} - \frac{i\bar{b}}{2\pi}\frac{z}{\bar{z}}\right). \tag{3}$$

Here, an overline denotes complex conjugation. Using Eq. (3), the strain fields are obtained according to, $\varepsilon \propto \frac{\partial u}{\partial z}, \frac{\partial u}{\partial \bar{z}}$. In the case of the infinite wall discussed in the main text, the resultant strain is obtained exactly by summing up a relevant series of pairs of dislocations at positions $z_n = i\frac{h}{2} + 2ihn, z'_n = -\frac{ih}{2} + 2ihn$ where $h$ is the distance between two dislocations with the same Burgers vector, and $b = 1 + i, b' = 1 - i$.

We calculate the diffuse scattering from the dislocation wall by a direct numerical evaluation of the kinetic scattering sum using the displacements in Eq. (3) on a square lattice. The results in Fig. 2c,d were obtained on a 400 × 400 lattice, using $|b| = |b'| = 2\sqrt{2}$ (assuming a dislocation bunching factor of two, i.e., pairs of dislocations) and nominal $h = 32$ lattice units. For the larger bunching factor calculations (Fig. S5), an 800 × 800 lattice was used, as well as $h = 64$ lattice units to obtain the same dislocation density per unit wall length. Scatterings from a horizontal and a vertical wall were summed incoherently (addition of intensities). For a more realistic comparison with experiment, we introduced noise by randomly shifting each dislocation by 0 or 1 lattice units along the wall direction. This also gives an average $h$ increased by 0.5 lattice units, decreases finite-size artifacts, and limits the length of the cross-shaped features in reciprocal space.

By substituting the position-dependent $\varepsilon_{ij}(z)$ in $\Lambda_{ij}$, we obtain a set of three frequencies, $\omega_i^2(z) = \omega_T^2 + \Lambda_{ii}(z)$, which in turn define an inhomogeneous dimensionless parameter $r_i(z) = \omega_i^2(z)/\omega_T^2$ that sets the distance to the putative ferroelectric QCP for a transverse mode polarized along the $i$ axis (with $r_i < 0$ denoting the ordered state and $r_i > 0$ denoting the disordered state). We note that the main difference between a single dislocation and a wall is that the strain decays polynomially for the former, $\varepsilon \sim \frac{|b|}{r}$, and exponentially for the latter, $\varepsilon \sim e^{-r/|b|}$. To produce Fig. 5, we plot $r_{\min} = \min r_i$. Recall that the dimensionless parameter $r = 1$ for bulk undeformed STO and $r = 0$ at the putative QCP. From DFT calculations[46], we estimate $\lambda_1 \approx -5 \times 10^3 \text{meV}^2$ and $\lambda_2 = -2 \times 10^4 \text{meV}^2$. We set the polar mode frequency to $\omega_T^2 = (2 \text{ meV})^2$ for definitiveness. Upon examining Eq. (2), it is clear that very close to the dislocation core, where the strain $\varepsilon \sim 1$, the physics is dominated by the strain terms, since $|\lambda_1|$ and $|\lambda_2|$ are very large compared to the soft mode. However, there will be a region around any dislocation where $\omega_{\min}$ will be reduced and thus $r_{\min}$ will become smaller than 1. In this region, $T_c$ will be enhanced, as proximity to the ferroelectric QCP is known, both experimentally[26,45] and theoretically[41,42,43], to enhance $T_c$. We did not attempt a quantitative estimate of the effect of the softening on $T_c$, as this is model dependent and not known at the QCP[41,42,43]. However, one can expect a significant change in $T_c$ if $r$ changes by a quantity of order one. Thus, the shaded regions in Fig. 6 correspond to $0 < r_{\min} \leq 1/2$.

We note that the real material will also have wall intersection points, which act like strain multipoles and should have an associated long-range strain component. However, from the fact that we do not observe appreciable Bragg peak broadening, we conclude that their volume fraction must be very small; this is in line with the TEM observation[16] of a characteristic dislocation bundle distance in the micrometre range.

**Acknowledgements**

We thank L. J. Thompson and Z. Jiang for help with sample preparation, S. L. Griffitt and A. Najev for assistance with the design and manufacturing of polishing rigs, C. N. Irfan Habeeb for help with figures, D. Robinson and S. Rosenkranz for assistance with X-ray scattering experiments, and B. I. Shklovskii, Y. Ayino, V. Pribiag, B. Kalisky and J. Ruhman for discussions and comments. The work at the University of Minnesota was funded by the U.S. Department of Energy through the University of Minnesota Center for Quantum Materials, under Grant No. DE-SC-0016371. The work at Argonne was supported by the U.S. Department of Energy, Office of Science, Basic Energy Sciences, Materials Sciences and Engineering Division. A portion of this research used resources at the Spallation Neutron Source, a DOE Office of Science User Facility operated by the Oak Ridge National Laboratory. This research used resources of the Advanced Photon Source, a U.S. Department of Energy (DOE) Office of Science User Facility operated for the DOE Office of Science by Argonne National Laboratory under Contract No. DE-AC02-06CH11357. DP acknowledges support from the Croatian Science Foundation through grant no. UIP-2020-02-9494. Sputtering and contacting of samples was conducted in the Minnesota Nano Center, which is supported by the National Science Foundation through the National Nano Coordinated Infrastructure Network, Award Number NNCI -1542202.

**Author Contributions**

DP and MG conceived the research; SH, RJS, DP, BD and JR performed transport and susceptibility measurements; SH, ZWA, DP and YL performed neutron scattering experiments and analysed data; ZWA, RJS, DP, MJK and RO performed X-ray scattering experiments and analysed data; LY and YL performed Raman scattering experiments; SH, LY, DP and YL analysed Raman data; AK and RMF performed calculations; CL provided and characterized samples; CL and DP oversaw transport measurements by SH, BD and JR; ML and DP designed and manufactured the pressure cells; DP, AK and MG wrote the paper, with input from all authors.

**Data and materials availability**

All data and materials are available from corresponding authors upon reasonable request.

and DP oversaw transport measurements by SH, BD and JR; ML and DP designed and manufactured the pressure cells; DP, AK and MG wrote the paper, with input from all authors.

**Data and materials availability**

All data and materials are available from corresponding authors upon reasonable request.

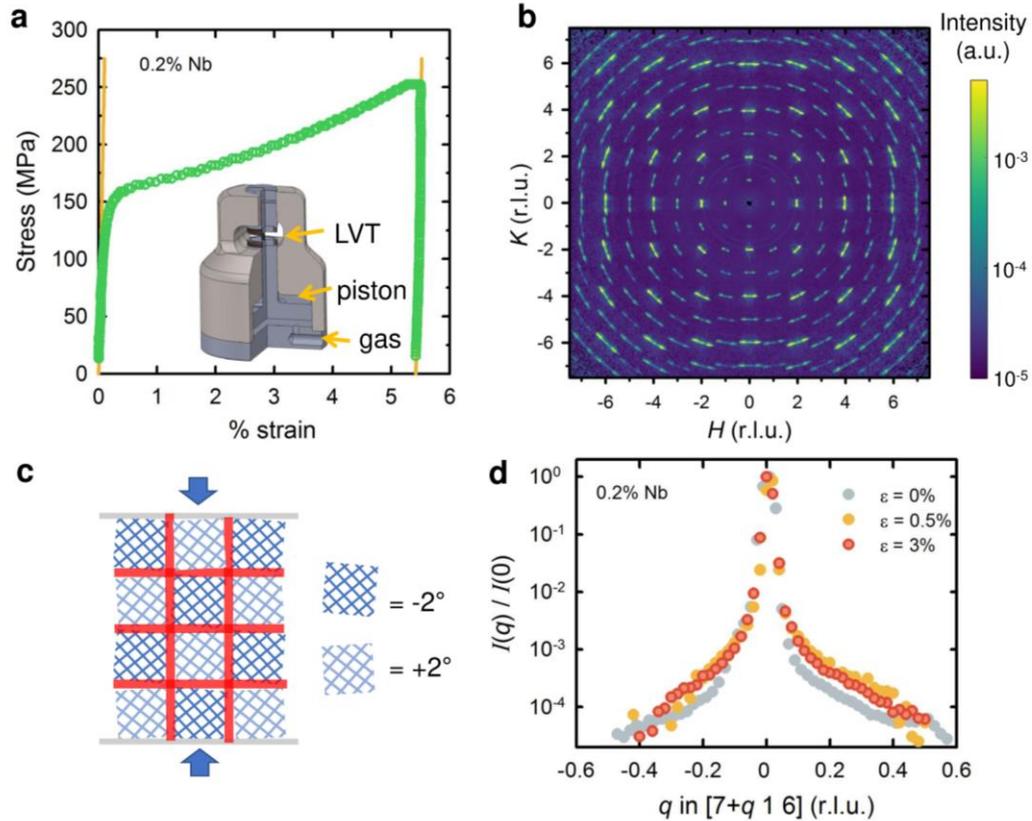

**Fig. 1 | Structure of plastically deformed STO.** (**a**) Representative engineering stress-strain curve for a 0.2% Nb-doped STO crystal at 300 K, with compressive uniaxial stress along one of the cubic principal axes. The orange lines indicate the elastic regime with Young's modulus ~270 GPa. An extended plastic region is seen, with work hardening at high strain, similar to undoped STO[16]. The inset is a schematic of the high-force uniaxial pressure cell, which uses a piston and gas pressure to apply stress on a polished sample. A linear variable transformer (LVT) is used to measure sample deformation. (**b**) Diffuse neutron scattering for an oxygen-vacancy-doped (OVD) STO sample with compressive deformation (engineering strain) ε = 4.2%, at 10 K. The sample was deformed *ex situ* at 300 K, with stress along [010]. The *HK*0 plane, which contains the stress direction, displays Bragg-peak elongations (asterisms) that are absent in other planes (Fig. S3). (**c**) Simplified cross-section schematic of a plastically deformed crystal that would show surface steps (Fig. S2) and asterisms. Crossed lines represent [110] and [1-10] crystallographic planes, which are the dominant dislocation slip planes. The differently shaded blue regions are homogeneously tilted by +2° and -2°, and red lines represent dislocation structures (walls) with high dislocation density. In a real material, there is a distribution of tilts and sizes of the regions. (**d**) Comparison of diffuse X-ray scattering profiles around the 716 Bragg peak for 0.2% Nb samples with ε = 0%, 0.5% and 3%, at 30 K. Stress is along [010], and the feature around $q = -0.1$ in the ε = 0% sample is from a detector artefact. The tails of the peaks are likely dominated by dislocation scattering[34], and show 2-3 times higher intensity in the deformed samples. The ε = 0.5% sample does not show asterisms, whereas the ε = 3% sample does (Fig. S4), but the scattering appears similar, suggesting that the dislocation density does not significantly change with increasing strain above ε = 0.5%.

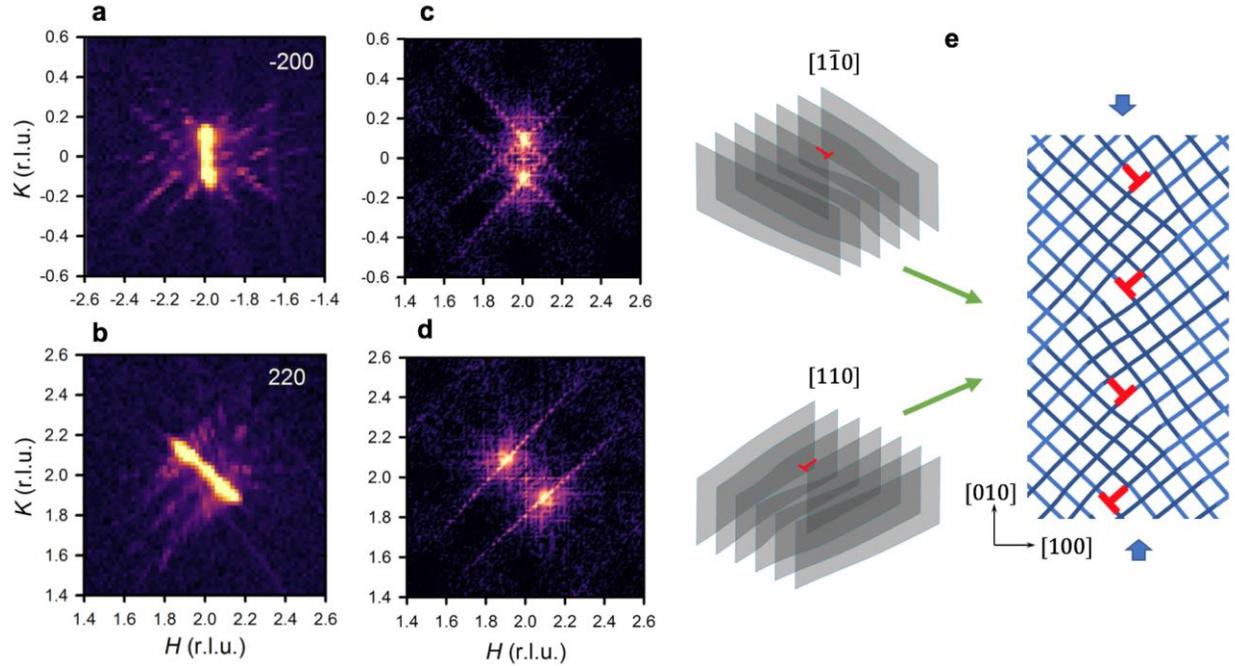

**Fig. 2 | Local structure of deformed STO from diffuse neutron scattering.** (**a**) Quasi-elastic diffuse neutron scattering around the -200 (**a**) and 220 (**b**) Bragg peaks of an OVD sample with ε = 4.2% (same sample as in Fig. 1b). The faint rings are background scattering from the sample holder. Several cross-like patterns are observed, with a grainy substructure, indicating long-range dislocation correlations. Other Brillouin zones are shown in Fig. S5. (**c**),(**d**) Calculated diffuse scattering from a dislocation wall of the form depicted in (**e**), with {110} slip planes. Scattering from horizontal and vertical domain walls is superimposed with equal weight. Streaks similar to the experimental features are found, and the correct Bragg peak splitting is obtained for a dislocation distance of 16 unit cells. The best agreement with experiment is obtained when each dislocation in (e) has a Burgers vector amplitude corresponding to two unit cells (see Methods and Fig. S6 for other values). The resolution-limited width of the diffuse streaks indicates a wall length of at least ~100 unit cells. Only two out of four possible slip planes are active (see Methods).

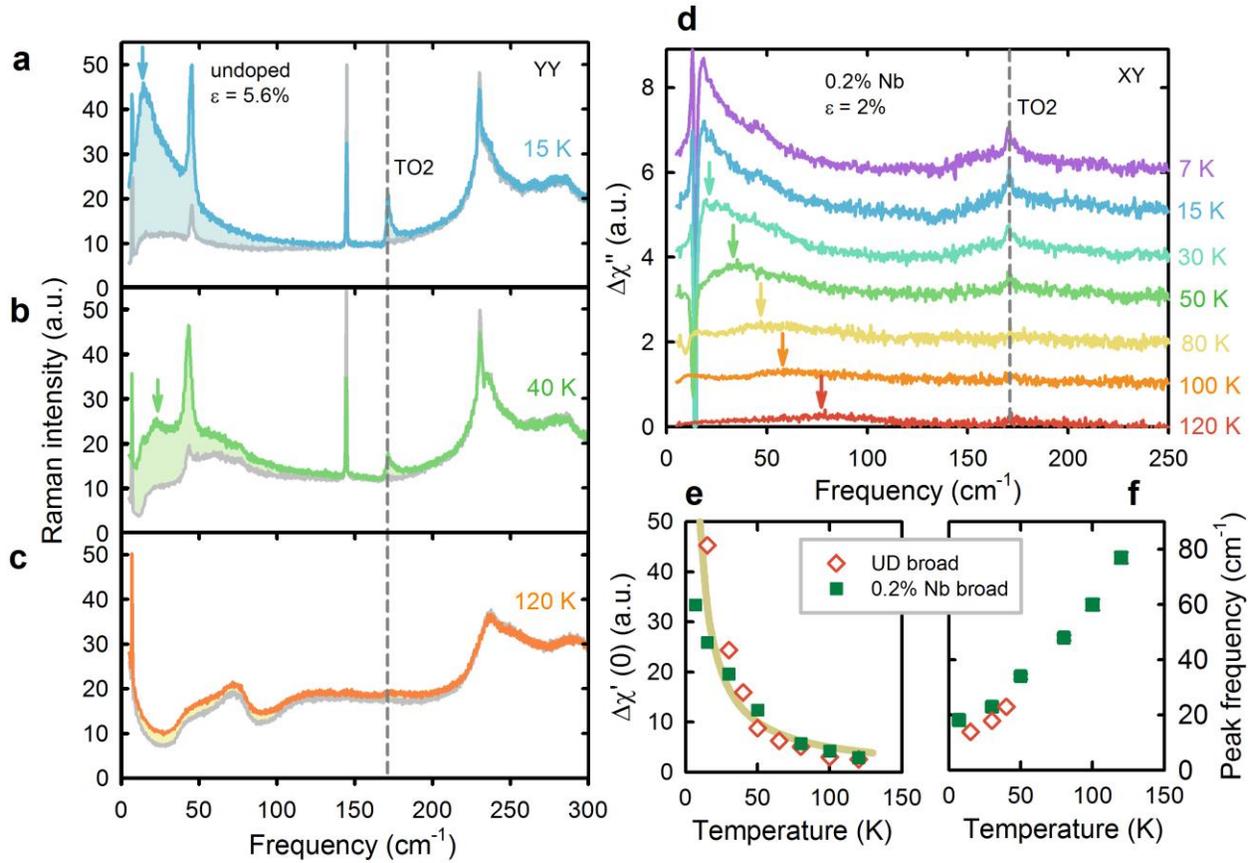

**Fig. 3 | Inversion-symmetry breaking and ferroelectric fluctuations in deformed STO.** (**a-c**) Polarized Raman scattering in two undoped STO crystals: undeformed (grey lines) and with ε = 5.6% deformation (coloured lines), measured at 15 K, 40 K and 120 K, respectively. Two deformation-induced features are seen: hard TO2 and TO4 modes indicative of inversion-symmetry breaking (see Fig. S7 for TO4), and a strong, broad feature at low energies that corresponds to polar fluctuations. (**d**) Difference between the imaginary parts of the Raman susceptibility, $\Delta\chi''$, of an undeformed and a deformed sample with 0.2% Nb reveals essentially the same features as in the undoped case (a-c). The sharp dip around 10 cm$^{-1}$ is an artifact of the subtraction of the signal from the undeformed sample. (**e**) Zero-frequency (static) real part of the Raman susceptibility obtained from the data in (a-d), showing the divergence of the broad low-energy feature upon cooling. The line is a $1/T$ Curie dependence, indicative of quantum critical fluctuations. (**f**) Characteristic frequency of the broad peak, marked by arrows in (a-d), as a function of temperature. A strong softening of the characteristic fluctuation scale is observed. Due to the phonon peak around 50 cm$^{-1}$, the broad peak position could not be unambiguously determined above 40 K for the undoped sample.

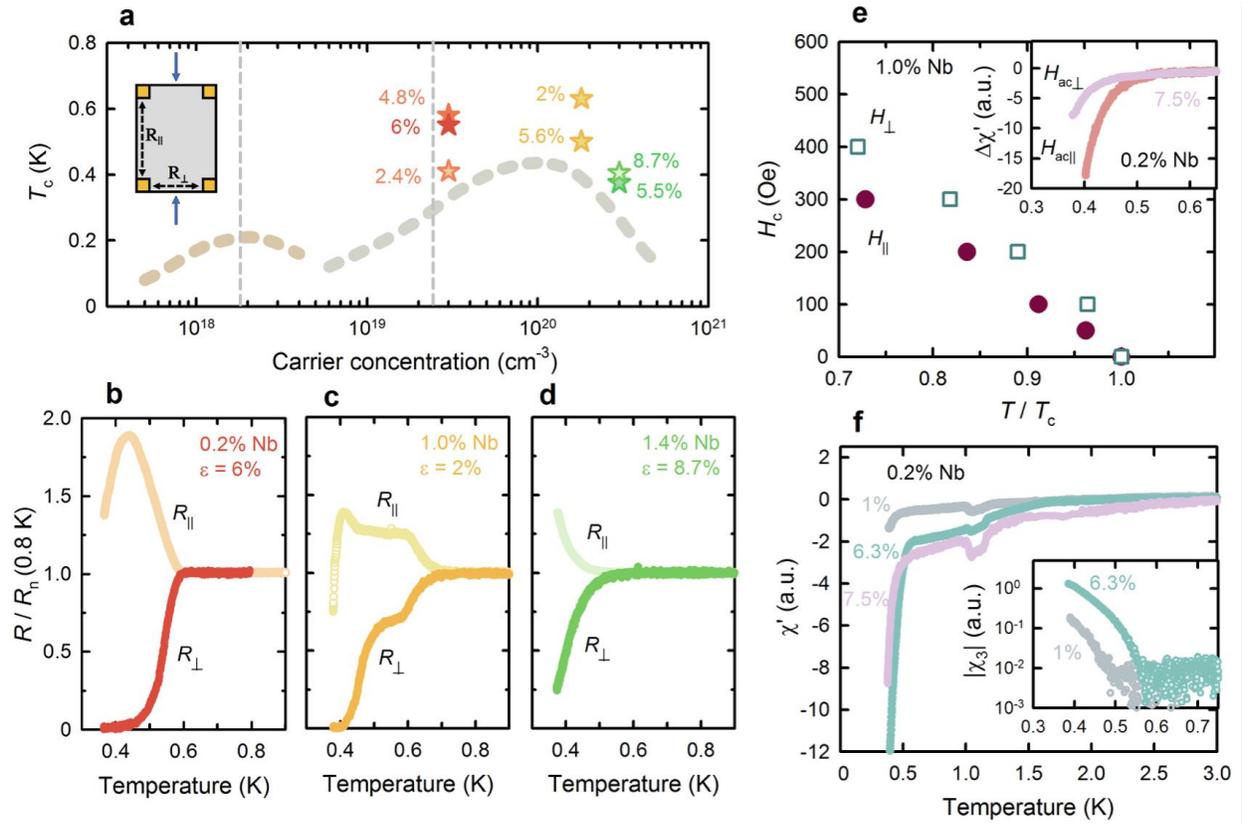

**Fig. 4 | Low-temperature superconducting properties of plastically deformed STO.** (**a**) Superconducting phase diagram of undeformed STO, showing the midpoints of resistive transitions for OVD (brown line) and Nb-doped (grey line) samples, based on [29]. Vertical lines: doping levels at which additional bands cross the Fermi level[29]. Stars: midpoints of the resistive transitions in the channel perpendicular to the stress direction ($R_\perp$, see sketch); colours correspond to those in (b-d) and indicate the deformation level. (**b**) Resistance for 0.2% Nb and $\varepsilon = 6\%$, measured parallel and perpendicular to the stress direction, normalized to the respective value ($R_n$) at 0.8 K. The bulk $T_c$ of undeformed STO at this doping level is about[12] 0.3 K, below the base temperature of our ³He refrigerator. (**c,d**) Same measurement as in (b) for 1% Nb and $\varepsilon = 2\%$, and 1.4% Nb and $\varepsilon = 8.7\%$, respectively. (**e**) Anisotropy of the superconducting response of a sample with 1% Nb and $\varepsilon = 5.6\%$: the critical magnetic field $H_c$, obtained from resistivity measurements, is substantially different for field orientations along and perpendicular to the stress direction. Inset: ac susceptibility of a 0.2% Nb, $\varepsilon = 7.5\%$ sample, with two orientations of the applied magnetic field, also showing a strong anisotropy. (**f**) Magnetic measurements of 0.2% Nb samples. Main panel: linear ac susceptibility at three deformation levels. The onset of significant diamagnetism corresponds to the resistivity downturn temperatures, and there is a small diamagnetic contribution at even higher temperatures. The feature around 1.2 K is due to the aluminium foil in the coil holder. Inset: third-order nonlinear susceptibility $\chi_3$ shows an onset temperature similar to that seen in the resistivity.

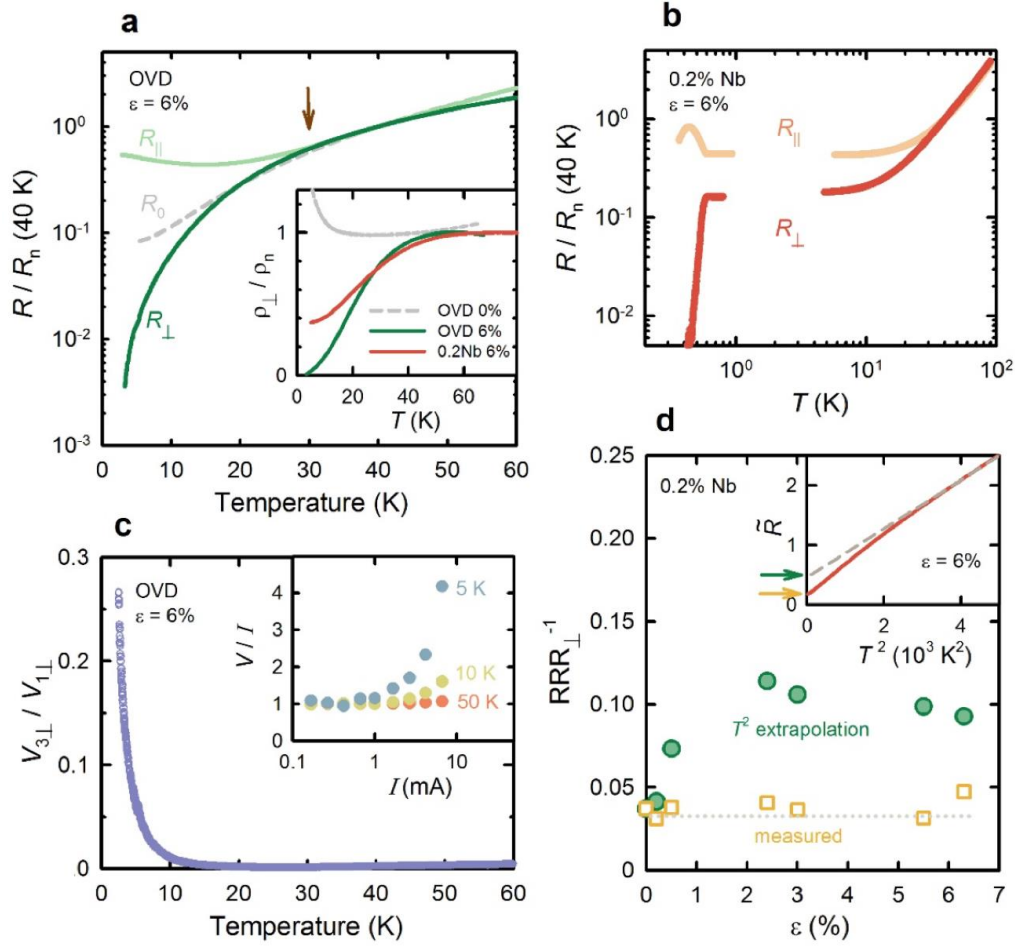

**Fig. 5 | Evidence for high-temperature superconducting correlations in plastically deformed STO.** (**a**) Resistivity of an OVD sample, with carrier density ~ $4 \cdot 10^{17}$ cm$^{-3}$ and $\varepsilon = 6\%$, measured parallel and perpendicular to deformation, normalized to values at 40 K. A dramatic anisotropy develops below ~30 K, similar to the superconducting behaviour in Fig. 3, and $R_\perp$ falls roughly an order of magnitude below the resistance of an undeformed sample ($R_0$, dashed line). Inset: ratio of measured $\rho_\perp$ to extrapolated normal-state resistivity $\rho_n = \rho_0 + a_2 T^2$, as observed for an undeformed OVD sample, an OVD sample with $\varepsilon = 6\%$, and a sample with 0.2% Nb and $\varepsilon = 6\%$ (**b**) $R_\perp$ and $R_\parallel$ for 0.2% Nb and $\varepsilon = 6\%$: a significant anisotropy develops already at temperatures much higher than the resistive $T_c$. (**c**) Third-order nonlinear response of an OVD sample with $\varepsilon = 6\%$ (same as in (a)), plotted as the ratio between third-harmonic and first-harmonic voltage. Inset: effective resistance (voltage-to-current ratio), normalized to the low-current values, versus applied current for the same sample. A strong nonlinearity develops at low temperatures, consistent with the third-harmonic measurement. (**d**) Inverse residual resistivity ratios (RRR$^{-1}$) defined as $R_\perp$(5 K)/$R_\perp$(90 K) for samples with 0.2% Nb, as a function of deformation. Open squares: measured RRR$^{-1}$ values. Circles: values extrapolated from a pure $T^2$ dependence. Inset: $T^2$-extrapolated (green arrow) and measured (orange arrow) residual resistivities, with data scaled to the respective values at 90 K ($\tilde{R}$), for a sample with $\varepsilon = 6\%$. All samples were cut from the same crystal before deformation.

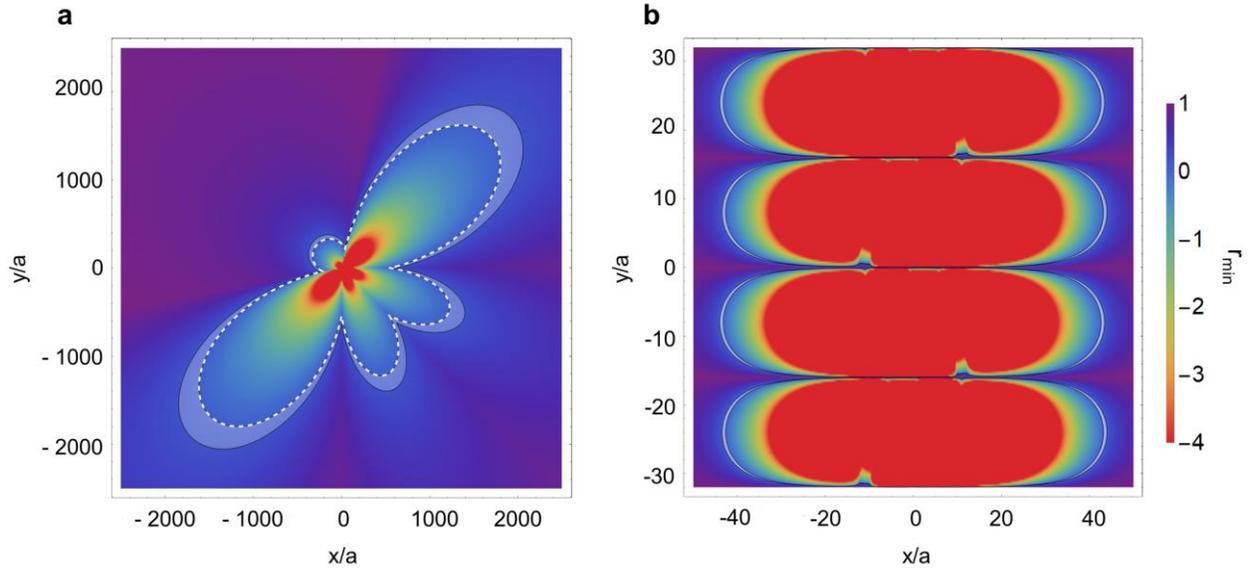

**Fig. 6 | Local $T_c$ enhancement due to ferroelectric fluctuations enhanced by dislocation-induced strain.** (**a**) Inhomogeneous dimensionless parameter $r_{\min} = \omega_{\min}^2/\omega_T^2$, which measures the distance to the ferroelectric phase transition. $r_{\min} = 1$: bulk undeformed sample. $r_{\min} < 0$: ferroelectric order. $r_{\min}(x, y)$ is calculated from the strain field generated by a single dislocation with Burgers vector along [110] and magnitude $b = a$, where $a = 3.9$ Å is the lattice constant. The white shaded area corresponds to $0 < r_{\min} < 1/2$, where an enhancement of $T_c$ of the order 1 is expected. For added clarity, the contour $r_{\min} = 0$ (where the system is at the QCP) is indicated as a thick black line. Deeper toward the center of the image, i.e., approaching the dislocation core, strain is so strong that ferroelectricity sets in. Outside the shaded region, i.e., farther from the dislocation core, strain is weak and $T_c$ enhancement is weak. (**b**) Similar $r_{\min}(x, y)$ map for a dislocation wall of the type shown in Fig. 2e. Note that here the width of the white area is comparable to $a$, a result of the rapid decay of the strain away from the wall. Whereas at small distances from the dislocation there is a microscopic substructure (the "blips" at x/a ~ ±15 essentially arise from the fact that $r_{\min}$ tracks the lowest of the three TO mode frequencies, which induces an effective nonlinearity), in the region where $T_c$ is enhanced, this structure is essentially gone, again due to the rapid decay. This implies that the $T_c$ enhancement is largely independent of the details of the wall structure. Note that the figure depicts a two-dimensional cut of the three-dimensional system, i.e., the shaded area is a cross section of a thin sheet of enhanced $T_c$.

# References

[1] Pfeiffer, E. R. & Schooley, J. F. Effect of stress on the superconductive transition temperature of strontium titanate. *Phys. Rev. Lett.* **19**, 783-785 (1967).

[2] Chu, J.-H. *et al.*, In-plane resistivity anisotropy in an underdoped iron arsenide superconductor. *Science* **329,** 824-826 (2010).

[3] Steppke, A. *et al*. Strong peak in $T_c$ of Sr$_2$RuO$_4$ under uniaxial pressure. *Science* **335**¸ eaaf9398 (2017).

[4] Herrera, C. *et al*. Strain-engineered interaction of quantum polar and superconducting phases. *Phys. Rev. Materials* **3,** 124801 (2019).

[5] Jakobsen, B. *et al*. Formation and subdivision of deformation structures during plastic deformation. *Science* **321**, 889-892 (2006).

[6] Sugiyama, I., Shibata, N., Wang, Z., Kobayashi, S. Yamamoto, T. & Ikuhara, Y. Ferromagnetic dislocations in antiferromagnetic NiO. *Nat. Nanotech.* **8**, 266-270 (2013).

[7] Shimada, T., Xu, T., Araki, Y. Wang, J. & Kitamura, T. Multiferroic dislocations in ferroelectric PbTiO$_3$. *Nano Lett.* **17**, 2674-2680 (2017).

[8] Kolhatkar, G. *et al*. Second harmonic generation investigation of symmetry breaking and flexoelectricity induced by nanoindentations in SrTiO$_3$. *Adv. Funct. Mater.* **29**, 1901266 (2019).

[9] Willa, R., Hecker, M., Fernandes, R. M. & Schmalian, J. Inhomogeneous time-reversal symmetry breaking in Sr$_2$RuO$_4$. *arxiv:2011.01941* (2020).

[10] Masuda, K., Lich, L. V., Shimada, T. & Kitamura, T. Periodically-arrayed ferroelectric nanostructures induced by dislocation structures in strontium titanate. *Phys. Chem. Chem. Phys.* **21**, 22756-22762 (2019).

[11] Schooley, J. F., Hosler, W. R. & Cohen, M. L. Superconductivity in semiconducting SrTiO$_3$. *Phys. Rev. Lett.* **12**, 474-475 (1964).

[12] Collignon, C., Lin, X., Rischau, C. W., Fauqué, B. & Behnia, K. Metallicity and superconductivity in doped strontium titanate. *Annu. Rev. Cond. Matt. Phys.* **10**, 25-44 (2019).

[13] Gastiasoro, M. N., Ruhman, J. & Fernandes, R. Superconductivity in dilute SrTiO$_3$: a review. *Ann. Phys.* **417**, 168107 (2020).

[14] Swartz, A. G. *et al*. Polaronic behaviour in a weak-coupling superconductor. *Proc. Nat. Acad. Sci. USA* **115**, 1475-1480 (2018).

[15] Cava, R. J. Oxide superconductors. *J. Am. Ceram. Soc.* **83**, 5-28 (2000).

[16] Gumbsch, P., Taeri-Baghbadrani, S., Brunner, D., Sigle, W. & Rühle, M. Plasticity and inverse brittle-to-ductile transition in strontium titanate. *Phys. Rev. Lett.* **87,** 085505 (2001).

[17] Hirel, P., Carrez, P. & Cordier, P. From glissile to sessile: effect of temperature on <110> dislocations in perovskite materials. *Scripta Materialia* **120**, 67-70 (2016).

[18] Gay, P. & Honeycombe, R. W. K. X-ray asterisms from deformed crystals. *Proc. Phys. Soc. A* **64**, 844-845 (1951).

[19] Lewis, D. X-ray microbeam study of deformation bands in aluminium. *Brit. J. Appl. Phys.* **11**, 162-164 (1960).

[20] Barabash, R. I. & Klimanek, P. X-ray scattering by crystals with local lattice rotation fields. *J. Appl. Cryst.* **32**, 1050-1059 (1999).

[21] Cowley, R. A. The phase transition of strontium titanate. *Phil. Trans. R. Soc. Lond.* A **354**, 2799-2814 (1996).

[22] Haeni, J. H. *et al*. Room-temperature ferroelectricity in strained $SrTiO_3$. *Nature* **430**, 758-761 (2004).

[23] Burke, W. J. & Pressley, R. J. Stress induced ferroelectricity in $SrTiO_3$. *Solid State Communications* **9**, 191 (1971).

[24] M. J. Coak, C. R. S. Haines, C. Liu, G. G. Guzmán-Verri, and S. S. Saxena. Pressure dependence of ferroelectric quantum critical fluctuations. *Phys. Rev. B.* **100**, 214111 (2019).

[25] Gao, P. *et al.*, Atomic-scale measurement of flexoelectric polarization at $SrTiO_3$ dislocations. *Phys. Rev. Lett.* **120**, 267601 (2018).

[26] Rischau, C. W. *et al*. A ferroelectric quantum phase transition inside the superconducting dome of $Sr_{1-x}Ca_xTiO_{3-\delta}$. *Nat. Phys.* **13**, 643-648 (2017).

[27] Gallais, Y. et al., Observation of incipient charge nematicity in $Ba(Fe_{1-x}Co_x)_2As_2$. *Phys. Rev. Lett.* **111**, 267001 (2013).

[28] Van Mechelen, J. L. M. et al. Electron-phonon interaction and charge carrier mass enhancement in $SrTiO_3$. *Phys. Rev. Lett.* **100**, 226403 (2008).

[29] Lin, X. *et al*. Critical doping for onset of two-band superconductivity in $SrTiO_{3-\delta}$. *Phys. Rev. Lett.* **112**, 207002 (2014).

[30] Ayino, Y., Yue, J., Wang, T., Jalan, B. & Pribiag, V. S. Evidence for multi-band superconductivity above the Pauli limit in $Nd_xSr_{1-x}TiO_3$. *arXiv:1812.02875* (2018).

[31] Liu, C. *et al*. Two-dimensional superconductivity and anisotropic transport at $KTaO_3$ (111) interfaces. *Science* **371**, 716-721 (2021).

[32] Vagli, R., Attanasio, C., Maritato, L. & Ruosi, A. Explanation of the resistance-peak anomaly in nonhomogeneous superconductors. *Phys. Rev. B* **47**, 15302-15303 (1993).

[33] Lin, X., Fauqué, B. & Behnia, K. Scalable $T^2$ resistivity in a small single-component Fermi surface. *Science* **349**, 945-948 (2015).

[34] Patterson, E. A. *et al*. Temperature-dependent deformation and dislocation density in $SrTiO_3$ (001) single crystals. *J. Am. Ceram. Soc.* **99**, 3411-3420 (2016).

[35] Spinelli, A., Torija, M. A., Liu, C., Jan, C. & Leighton, C. Electronic transport in doped $SrTiO_3$: conduction mechanisms and potential applications. *Phys. Rev. B* **81**, 155110 (2010).

[36] Coey, J. M. D., Venkatesan, M. & Stamenov, P. J. Surface magnetism of strontium titanate. *J. Phys.: Condens. Matter* **28**, 485001 (2016).

[37] Aird, A. & Salje, E. K. H. Sheet superconductivity in twin walls: experimental evidence of $WO_{3-x}$. *J. Phys. Condens. Matter* **10**, L377-L380 (1998).

38 Lin, Y.-H., Chen, Y. & Goldman, A. M. Indications of superconductivity at somewhat elevated temperatures in strontium titanate subjected to high electric fields. *Phys. Rev. B* **82**, 172507 (2010).

39 Uwe, H. & Sakudo, T. Stress-induced ferroelectricity and soft phonon modes in $SrTiO_3$. *Phys. Rev. B* **13**, 271-286 (1976).

40 Rowley, S. *et al*. Superconductivity in the vicinity of a ferroelectric quantum phase transition. *arXiv:1801.08121* (2018).

41 Edge, J. M., Kedem, Y., Aschauer, U., Spaldin, N. A. & Balatsky. A. V. Quantum critical origin of the superconducting dome in $SrTiO_3$. *Phys. Rev. Lett.* **115**, 247002 (2015).

42 Kozii, V., Bi, Z. & Ruhman, J. Superconductivity near a ferroelectric quantum critical point in ultralow-density Dirac materials. *Phys. Rev. X* **9**, 031046 (2019).

43 Gastiasoro, M. N., Trevisan, T. V. & Fernandes, R. M. Anisotropic superconductivity mediated by ferroelectric fluctuations in cubic systems with spin-orbit coupling. *Phys. Rev. B* **101**, 174501 (2020).

44 van der Marel, D., Barantani, F. & Rischau, C. W. Possible mechanism for superconductivity in doped $SrTiO_3$. *Phys. Rev. Research* **1**, 013003 (2019).

45 Ahadi, K. *et al*. Enhancing superconductivity in $SrTiO_3$ films with strain. *Sci. Adv.* **5**, eaaw0120 (2019).

46 K. Dunnett, A. Narayan, N. A. Spaldin, and A. V. Balatsky. Strain and ferroelectric soft mode induced superconductivity in strontium titanate. *Phys. Rev. B* **97**, 144506 (2018).

47 Pelc, D., Anderson, Z., Yu, B., Leighton, C. & Greven, M. Universal superconducting precursor in three classes of unconventional superconductors. *Nat. Commun.* **10**, 2729 (2019).

48 Pai, Y.-Y. *et al*. One-dimensional nature of superconductivity at the $LaAlO_3$/$SrTiO_3$ interface *Phys. Rev. Lett.* **120**, 147001 (2018).

49 McCalla, E., Gastiasoro, M. N., Cassuto, G., Fernandes, R. M. & Leighton, C. Low-temperature specific heat of doped $SrTiO_3$: doping dependence of the effective mass and Kadowaki-Woods scaling violation. *Phys. Rev. Materials* **3**, 022001(R) (2019).

50 Jiang, M. P. *et al*. The origin of incipient ferroelectricity in lead telluride. *Nat. Commun.* **7**, 12291 (2016).

51 Aktas, O., Crossley, S., Carpenter, M. A. & Salje, E. K. H. Polar correlations and defect-induced ferroelectricity in cryogenic $KTaO_3$. *Phys. Rev. B* **90**, 165309 (2014).

52 Sleight, A. W. Bismuthates: $BaBiO_3$ and related superconducting phases. *Physica C* **514**, 152-165 (2015).

53 Schilling, J. S., High pressure effects, *in* Handbook of High-temperature Superconductivity, eds. J. R. Schrieffer, J. S. Brooks (Springer, Berlin, 2007).

54 Ye, F., Liu, Y., Whitfield, R., Osborn, R. & Rosenkranz, S. Implementation of cross correlation for energy discrimination on the time-of-flight spectrometer CORELLI. *J. Appl. Cryst.* **51**, 315-322 (2018).
29

[55] Michels-Clark, T. M., Savici, A. T., Lynch, V. E., Wang, X. & Hoffmann, C. M. Expanding Lorentz and spectrum corrections to large volumes of reciprocal space for single-crystal time-of-flight neutron diffraction. *J. Appl. Cryst.* **49**, 497-506 (2016).

[56] Krogstad, M. J. *et al*. Reciprocal space imaging of ionic correlations in intercalation compounds. *Nat. Mater.* **19,** 63–68 (2020).

[57] Drobac, Đ., Marohnić, Ž., Živković, I. & Prester, M. The role of lock-in phase setting in ac susceptibility measurements. *Rev. Sci. Instrum.* **84**, 054708 (2013).

[58] Landau, L. D. & Lifshitz, E. M. Theory of elasticity (Butterworth-Heinemann, Oxford, 1986)

Supplementary Material for

# Ferroelectric quantum criticality and enhanced superconductivity in plastically deformed strontium titanate

S. Hameed, D. Pelc, Z. W. Anderson, A. Klein, R. J. Spieker, L. Yue, B. Das, J. Ramberger, M. Lukas, Y. Liu, M. J. Krogstad, R. Osborn, Y. Li, C. Leighton, R. M. Fernandes, and M. Greven

## Supplementary Text

*Dislocation scattering.* The observed resistivity anisotropy that sets in below 30-50 K in deformed samples is striking, especially in OVD crystals (Fig. 5a). However, it is important to discuss its possible origins in dislocation scattering mechanisms that have been known to be active in other semiconductors such as GaN [1,2]. We argue here that the dislocations induced by plastic deformation cause a simple residual resistivity increase that is roughly the same for current flow along and perpendicular to the stress direction. Experimentally, this is seen from Fig. S9b: the two resistance channels $R_\parallel$ and $R_\perp$ for a sample with 0.2% Nb and $\varepsilon = 6\%$ show similar residual resistivities extrapolated from the $T^2$ temperature dependence. We note that a similar analysis is not feasible for OVD samples with very low carrier density, given the limited temperature range of the $T^2$ dependence (especially in $R_\parallel$), but Fig. S9a nevertheless indicates that an increase in dislocation density also influences the residual resistivity there. In semiconductors such as GaN, charged-dislocation scattering is also visible in the temperature dependence of the Hall number [2]. Yet the Hall number of deformed STO is temperature-independent and does not depend on the current direction (Fig. S12). It therefore is unlikely that the nontrivial temperature dependence of $R_\perp$ results from dislocation-induced changes of carrier density.

Anisotropic mobility is in principle expected for scattering from a single dislocation in a metal [3]. However, the slip planes in STO are {110}, and multiple slip systems are active for stress along [100]. This likely averages out the anisotropy and leads to a simple homogeneous residual resistivity increase. It is not clear how dislocation scattering could lead to a *decrease* of $R_\perp$ with deformation, as seen most clearly in OVD samples (Figs. 5a and S9a). The weak dependence of the measured RRR in the perpendicular channel on strain (Fig. 5d) is thus probably accidental and originates from the compensation of two effects: an increase of the extrapolated residual resistivity due to dislocations, and a decrease due to possible superconducting correlations that do not percolate.

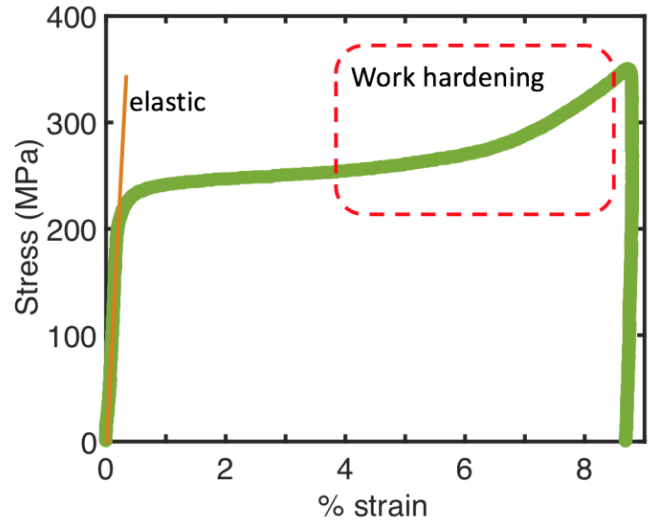

**Fig. S1 | Stress-strain diagram of doped strontium titanate.** Engineering stress-strain curve for a 2% Nb-doped STO crystal at 300 K, compressed along [010] to a high strain of ~8.7%. The elastic and work-hardening regimes are indicated.

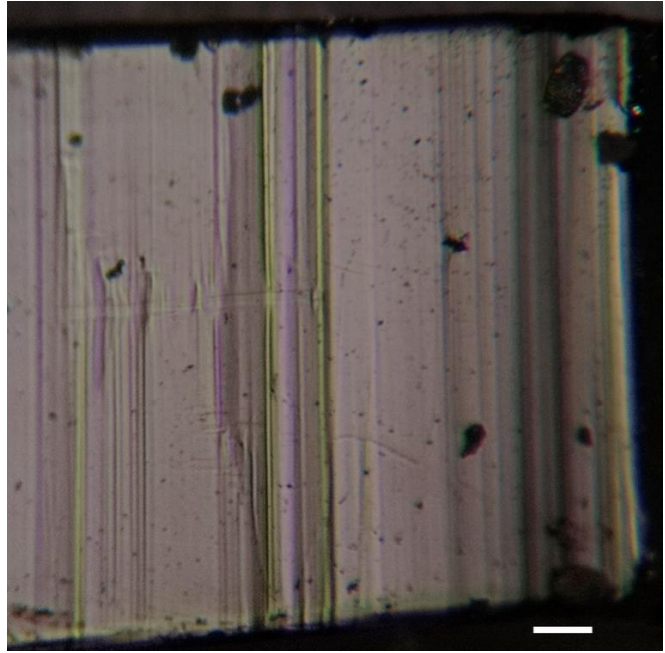

**Fig. S2 | Stress-induced surface deformation.** Photograph of the surface of a Nb-doped sample deformed to 7% in compression (the applied force direction is horizontal and along [010]), showing lines/ridges in a direction perpendicular to the stress. The incident angle of the light is roughly 45° with respect to the surface, and the scale bar is 100 μm.

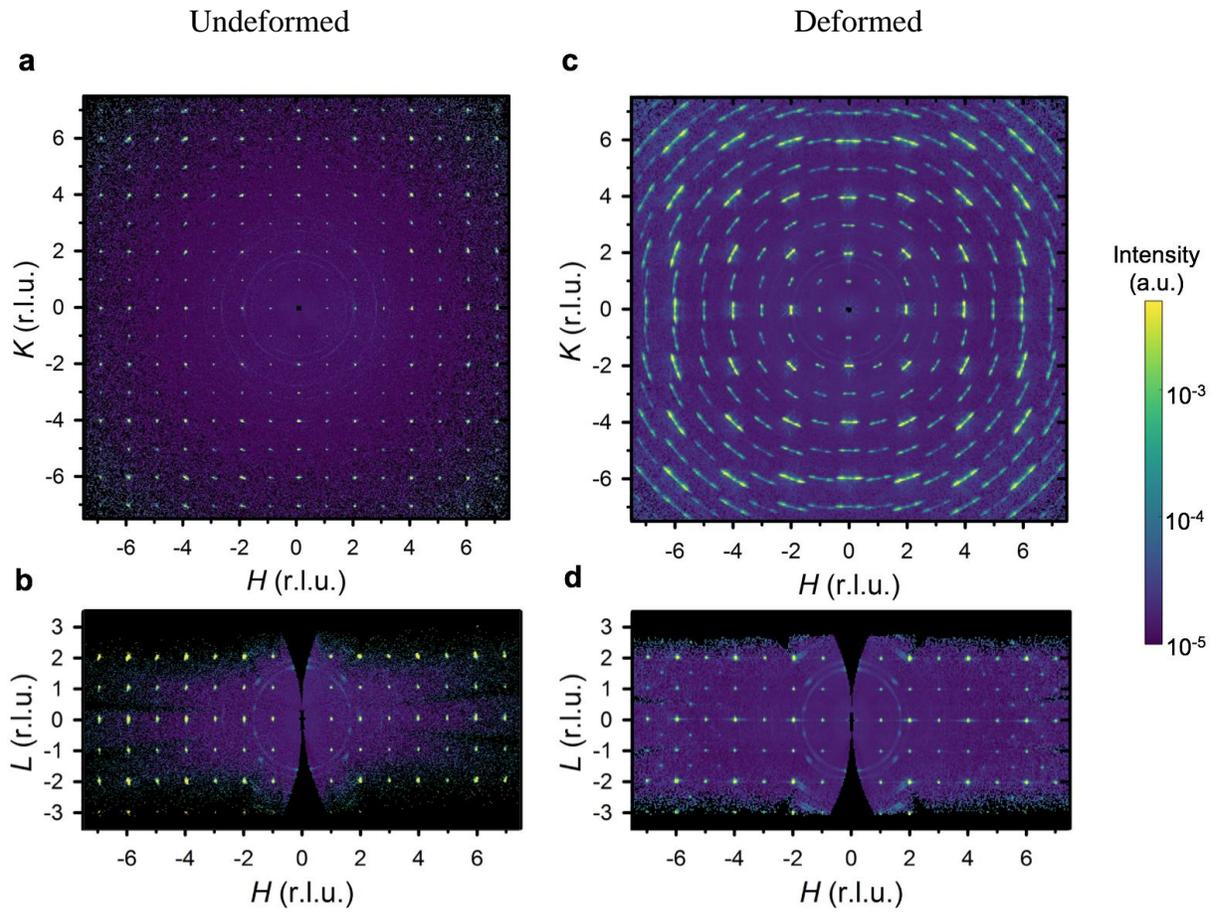

**Fig. S3 | Neutron scattering in undeformed and deformed STO.** Reciprocal space maps of (**a,b**) OVD undeformed and (**c,d**) 4.2% deformed STO single crystals in the *HK*0 and *H0L* planes, measured at 10 K. Sharp Bragg peaks and a slight mosaicity (visible in high Brillouin zones) are observed in the undeformed sample. For the deformed sample, asterisms (same as Fig. 1b) are seen in the *HK*0 plane which contains the stress direction. In contrast, the Bragg peaks remain sharp in other planes, as exemplified in (d) for *H0L*. Note that the usual superstructure peaks at half-integer positions are observed below the cubic-to-tetragonal phase transition of ~100 K, and that they also show asterisms whose intersections with the *H0L* plane are visible in higher zones in (d).

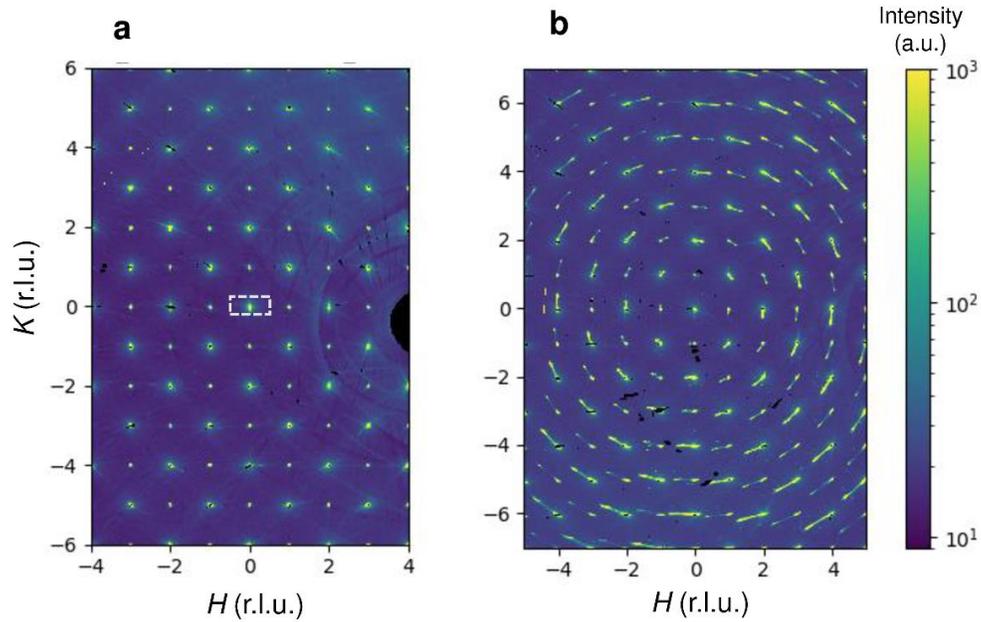

**Fig. S4 | Diffuse X-ray scattering in deformed STO.** Diffuse X-ray scattering at 30 K for samples with 0.2% Nb and (**a**) $\varepsilon = 0.5\%$ and (**b**) $\varepsilon = 3\%$. The $HK6$ plane is shown, and the white box in (a) marks the integration region used to obtain Fig. 1d. The black areas are masked detector artifacts (see Methods). Strong asterisms similar to the OVD sample with $\varepsilon = 4.2\%$ (Fig. 1b) are observed for the sample with high deformation in (b), while the sample with low deformation shows almost no asterisms.

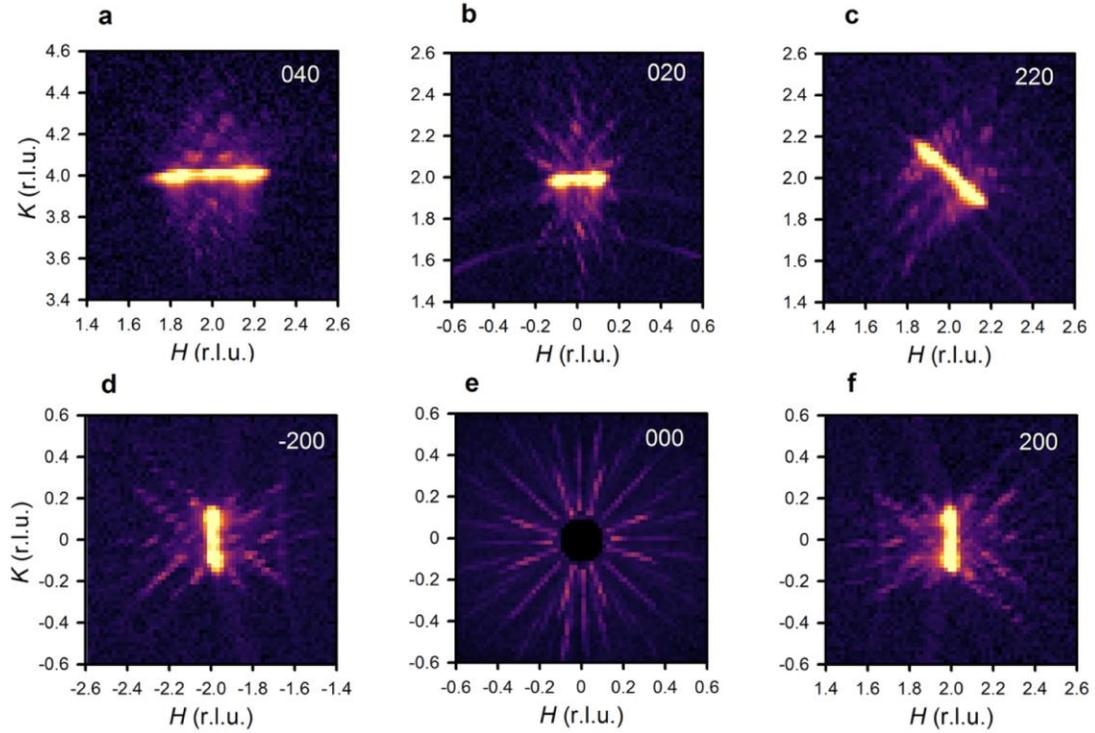

**Fig. S5 | Diffuse neutron scattering in deformed STO**. Bragg peaks and diffuse scattering in an OVD sample with ε = 4.2% (same as in Fig. 3), in different Brillouin zones near the: (**a**) 040 Bragg peak, (**b**) 020 Bragg peak, (**c**) 220 Bragg peak, (**d**) -200 Bragg peak, (**e**) reciprocal space origin, and (**f**) 220 Bragg peak (same as Fig. 3b). Cross-like patterns (highlighted in Fig. 3b) are most clearly seen in (d) and (f), with additional horizontal lines through the centres of the asterisms; the 220 zone shows diagonal streaks, while secondary diffuse peaks are visible in the higher-order 040 zone. The streaks around the origin in (**e**) are most likely from double Bragg scattering, but could include other contributions from long-range correlated dislocation structure.

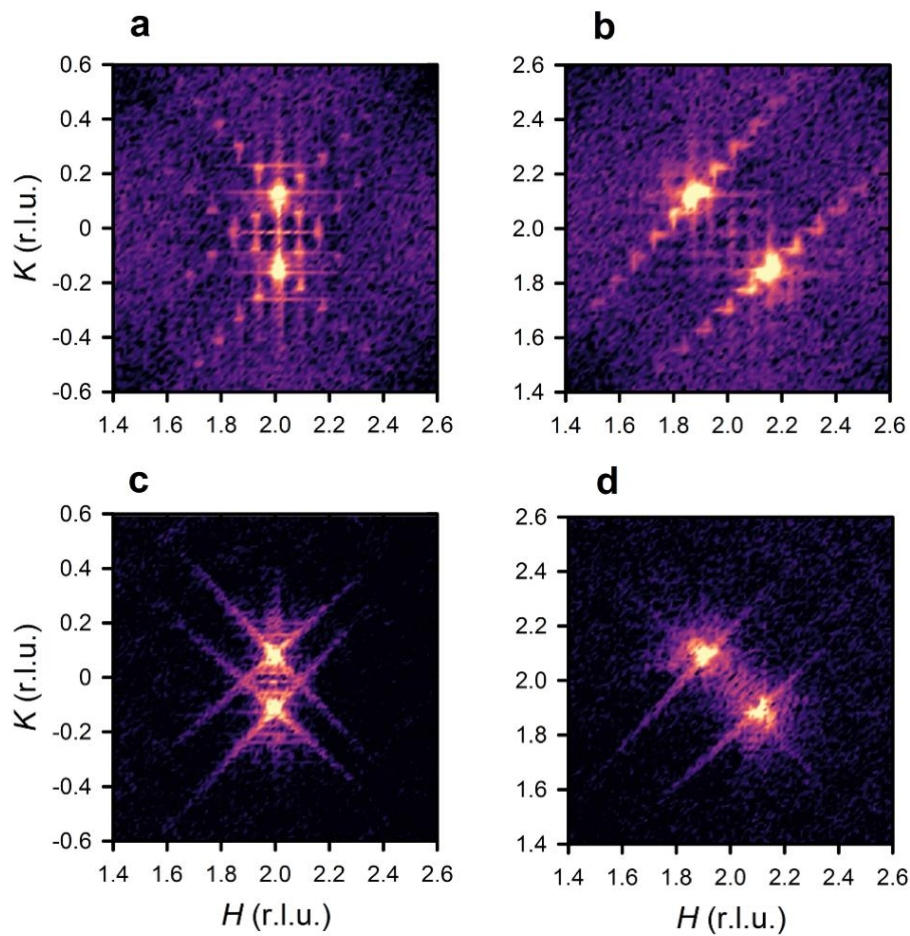

**Fig. S6 | Diffuse scattering from a dislocation wall. (a,b)** Calculated scattering from a dislocation wall as depicted in Fig. 2e, but without dislocation bunching (single dislocations), around the -200 and 220 Bragg positions, respectively. **(c,d)** Calculated scattering from a dislocation wall with four dislocations per bunch, same Bragg positions as above. The bunching clearly influences the periodicity of the diffuse scattering streak structures, and walls with significant bunching also show a sizable intensity on the line that joins the two split Bragg peaks.

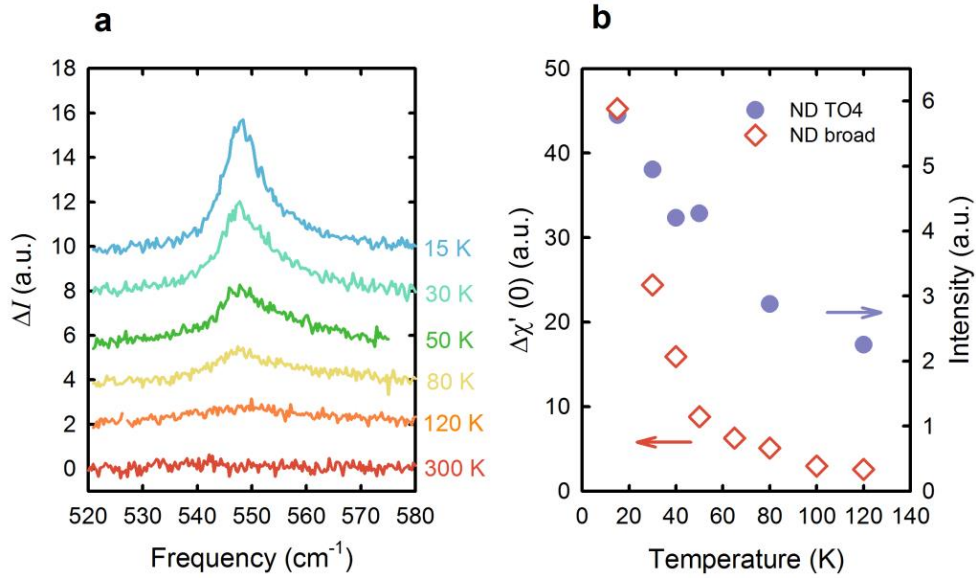

**Fig. S7 | Hard polar mode in deformed STO.** (**a**) Raman intensity difference between two undoped STO samples, with $\varepsilon = 0\%$ and $\varepsilon = 2\%$. The hard polar TO4 mode appears in the deformed sample, giving direct evidence of local inversion-symmetry breaking, similar to the TO2 mode in Fig. 4. (**b**) comparison of the integrated intensities of the TO4 mode and low-energy broad feature deformed undoped STO. The TO4 intensity clearly decays much slower compared to the Curie-Weiss-like dependence of the broad feature.

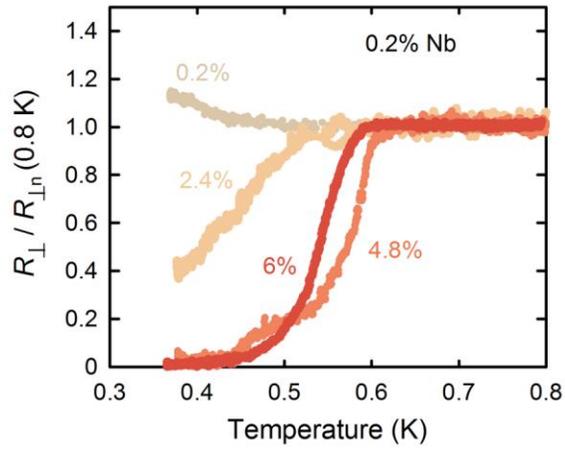

**Fig. S8 | Strain dependence of $T_c$ in a Nb-doped STO crystal.** Resistance data in the perpendicular channel for samples with 0.2% Nb and different deformation levels. A strong enhancement of the resistive $T_c$ with increasing deformation is observed.

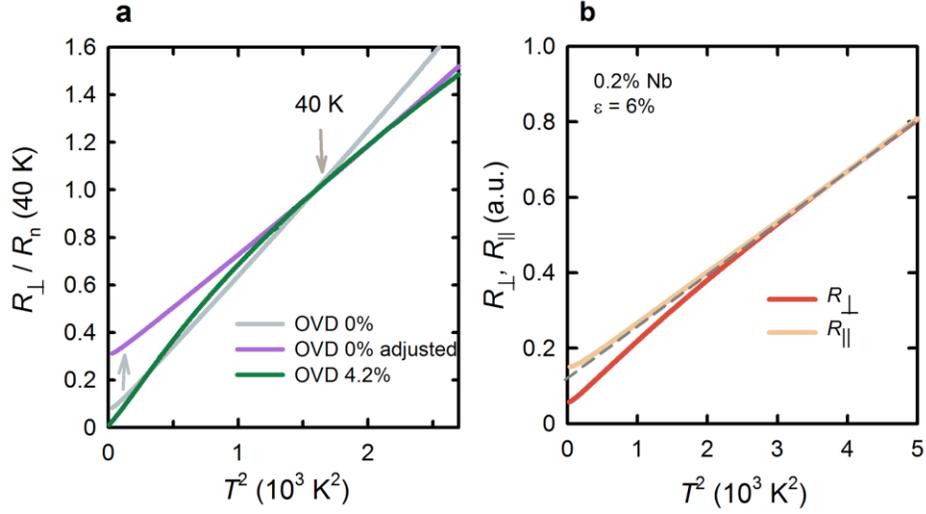

**Fig. S9 | Change of residual resistivity with deformation**. (**a**) Comparison of resistivities of deformed and undeformed OVD samples. The full green line is the resistivity perpendicular to the stress in a 4.2% deformed crystal, normalized to the value at 40 K; the grey line is the resistivity of an undeformed crystal annealed under the same conditions and normalized in the same way. A more meaningful comparison is obtained if we adjust for the fact that the residual resistivity at $T = 0$ increases with deformation (purple line). To do this, we shift the undeformed sample resistivity up by a constant offset and subsequently normalize the shifted curve to the value at 40 K, so as to match the slopes of the $T^2$ dependences in the two samples (in the limited temperature range where this is possible). In this way, the extrapolated normal-state resistivities overlap. (**b**) Resistivities parallel and perpendicular to the stress direction for a sample with 0.2% Nb and $\varepsilon = 6\%$ (same data as in Fig. 5b, normalized in such a way that they overlap within the $T^2$ regime above ~50 K). The residual resistivities extrapolated from a $T^2$ temperature dependence (dashed line) are similar for both directions, indicating that the increase in dislocation density upon plastic deformation leads to a simple direction-independent residual resistivity increase.

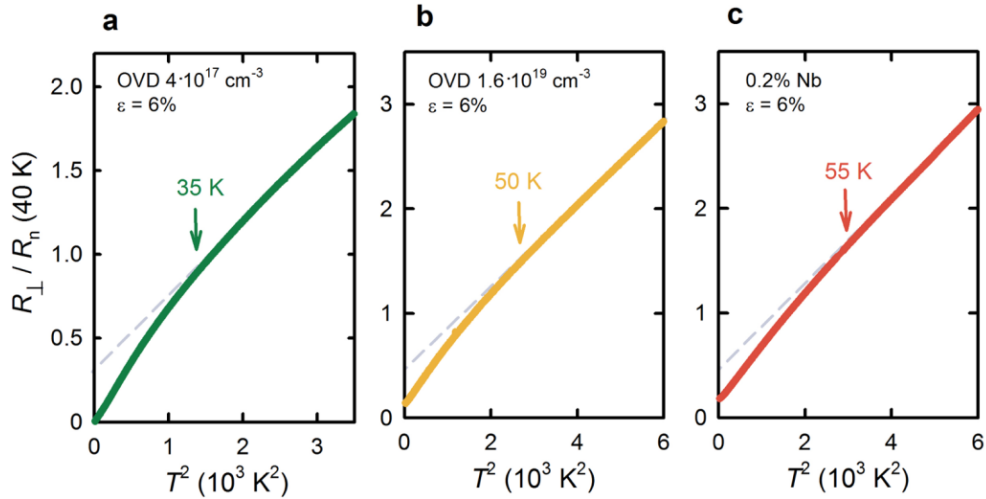

**Fig. S10 | Resistivity downturns in deformed STO in dependence on doping.** Normalized perpendicular resistances of two oxygen-vacancy-doped samples ((**a**) and (**b**)) and a sample with 0.2% Nb (**c**) plotted versus $T^2$. All samples are deformed to 6% in compression. The sample in (b) is the same as in (a), additionally annealed at 900°C for 1 h. A clear downward deviation in the resistance develops below the temperatures marked by arrows. Carrier densities are from Hall number measurements at 50 K. The samples in (**b**) and (**c**) have similar carrier densities, which indicates that the effect does not substantially depend on the type of dopant.

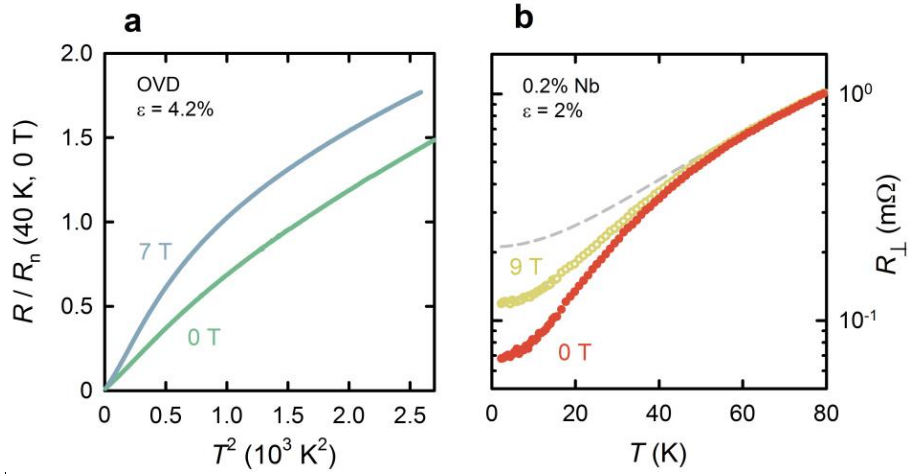

**Fig. S11 | Magnetic field effect on the perpendicular resistance**. (**a**) OVD sample, $\varepsilon = 4.2\%$, and (**b**) 0.2% Nb sample, $\varepsilon = 2\%$. Due to the low carrier concentration, the OVD sample has a large positive normal-state magnetoresistance, which enhances the relative resistivity decrease on cooling; the resistance still reaches very low values in a 7 T field. For the Nb-doped sample, the expected low-temperature magnetoresistance is ~20% at this doping level and field[4,5], whereas the observed value is roughly 100%. The dashed line is an extrapolation of the high-temperature $T^2$ behavior. The magnetic field is applied parallel to the shortest sample axis, perpendicular to both the stress and current directions.

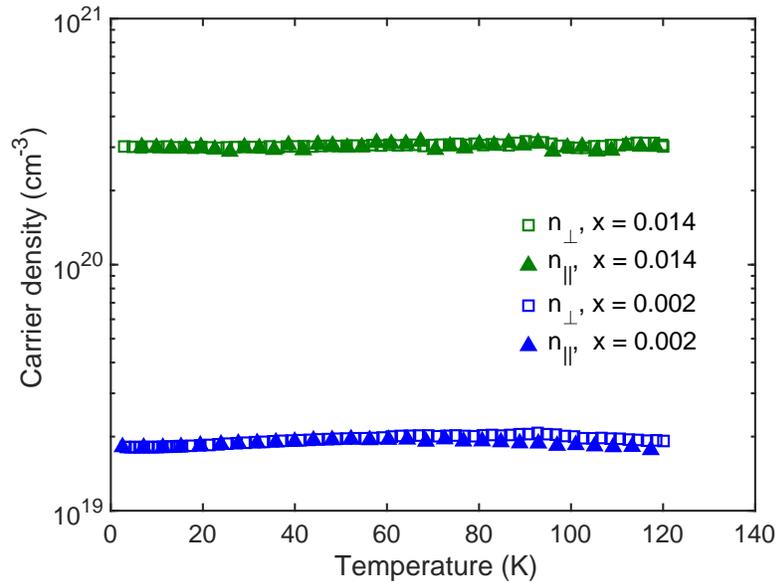

**Fig. S12 | Carrier density in deformed STO samples.** Electron density derived from Hall effect measurements in a 9 T field, for samples with 0.2% Nb, $\varepsilon = 2\%$ (blue symbols) and 1.4% Nb, $\varepsilon = 2\%$ (green symbols). There is no strong temperature dependence in either the direction parallel or perpendicular to stress. The density for the 0.2% Nb sample is about 0.7 times the expected density if every Nb dopant is substitutional and ionized. This difference is most likely due to acceptor impurities, and it is consistent with previous measurements on similar samples[5]. The density for the 1.4% Nb sample is slightly higher than expected, which could be related to current contact size corrections in Hall voltage measurements.

**Supplementary References**

[1] Weimann, N. G. & Eastman, L. F. Scattering of electrons at threading dislocations in GaN. *J. Appl. Phys.* **83**, 3656 (1998)

[2] Look, D. C. & Sizelove, J. R. Dislocation scattering in GaN. *Phys. Rev. Lett.* **82**, 1237 (1999)

[3] Dexter, D. L. Scattering of electrons in metals by dislocations. *Phys. Rev.* **86**, 770 (1952)

[4] Lin, X. *et al*. Critical doping for onset of two-band superconductivity in $SrTiO_{3-\delta}$. *Phys. Rev. Lett.* **112**, 207002 (2014)

[5] Spinelli, A., Torija, M. A., Liu, C., Jan, C. & Leighton, C. Electronic transport in doped $SrTiO_3$: conduction mechanisms and potential applications. *Phys. Rev. B* **81**, 155110 (2010)